\documentclass[%
 reprint,
superscriptaddress,
 amsmath,amssymb,
 aps,
]{revtex4-2}

\usepackage{graphicx}
\usepackage{dcolumn}
\usepackage{bm}
\usepackage{siunitx}
\usepackage{verbatim}
\usepackage[version=4]{mhchem}
\usepackage{pifont}
\usepackage{mathtools}

\DeclarePairedDelimiter\ket{\lvert}{\rangle}
\DeclarePairedDelimiterX\braket[2]{\langle}{\rangle}{#1\,\delimsize\vert\,\mathopen{}#2}

\begin{document}

\preprint{APS/123-QED}

\title{Phonon-Mediated Third-Harmonic Generation in Diamond}

\author{Jiaoyang Zheng}
 \affiliation{School of Applied and Engineering Physics, Cornell University, Ithaca, NY 14853, USA}
\author{Guru Khalsa}%
\affiliation{Department of Materials Science and Engineering, Cornell University, Ithaca, NY 14853, USA}%
\affiliation{Department of Physics, University of North Texas, Denton, TX 76203, USA}%

\author{Jeffrey Moses}
\affiliation{School of Applied and Engineering Physics, Cornell University, Ithaca, NY 14853, USA}%

\date{\today}

\begin{abstract}

We observe strongly anisotropic third-harmonic generation mediated by resonant sum-frequency driving of Raman phonons with THz light, extending light-induced dual control of structural and optical properties in solids. Either strong enhancement or strong suppression of the third harmonic covering six orders of magnitude can be achieved, a result of interference between purely electronic and phonon-mediated contributions to the polarization field. These findings enrich capabilities for tailoring nonlinear optics via phononics and for the spectroscopy of crystalline structural dynamics.

\end{abstract}

\maketitle


The close connection between structure, symmetry, and function is a principal driver of the search for new quantum materials. Leveraging this connection, prospects for ultrafast control of quantum materials have emerged employing the direct excitation of the crystalline lattice with light. 
Already, striking changes to functional properties such as superconductivity, magnetism and ferroelectricity have been achieved through the strong coupling between Raman-active phonons and infrared-active (IR-active) phonons driven by resonant THz laser pulses within the anharmonic regime \cite{disa2021engineering}. Alongside the manipulation of crystal structure, new research has focused on imparting giant changes to the optical properties of crystals through resonant IR-phonon excitation, mediated by the anharmonic lattice potential or nonlinear ionic displacement polarizability \cite{dolgaleva2015prediction, khalsa2021ultrafast, ginsberg2023phonon, zibod2023strong}.

\begin{figure}
\includegraphics[width=8.3 cm]
{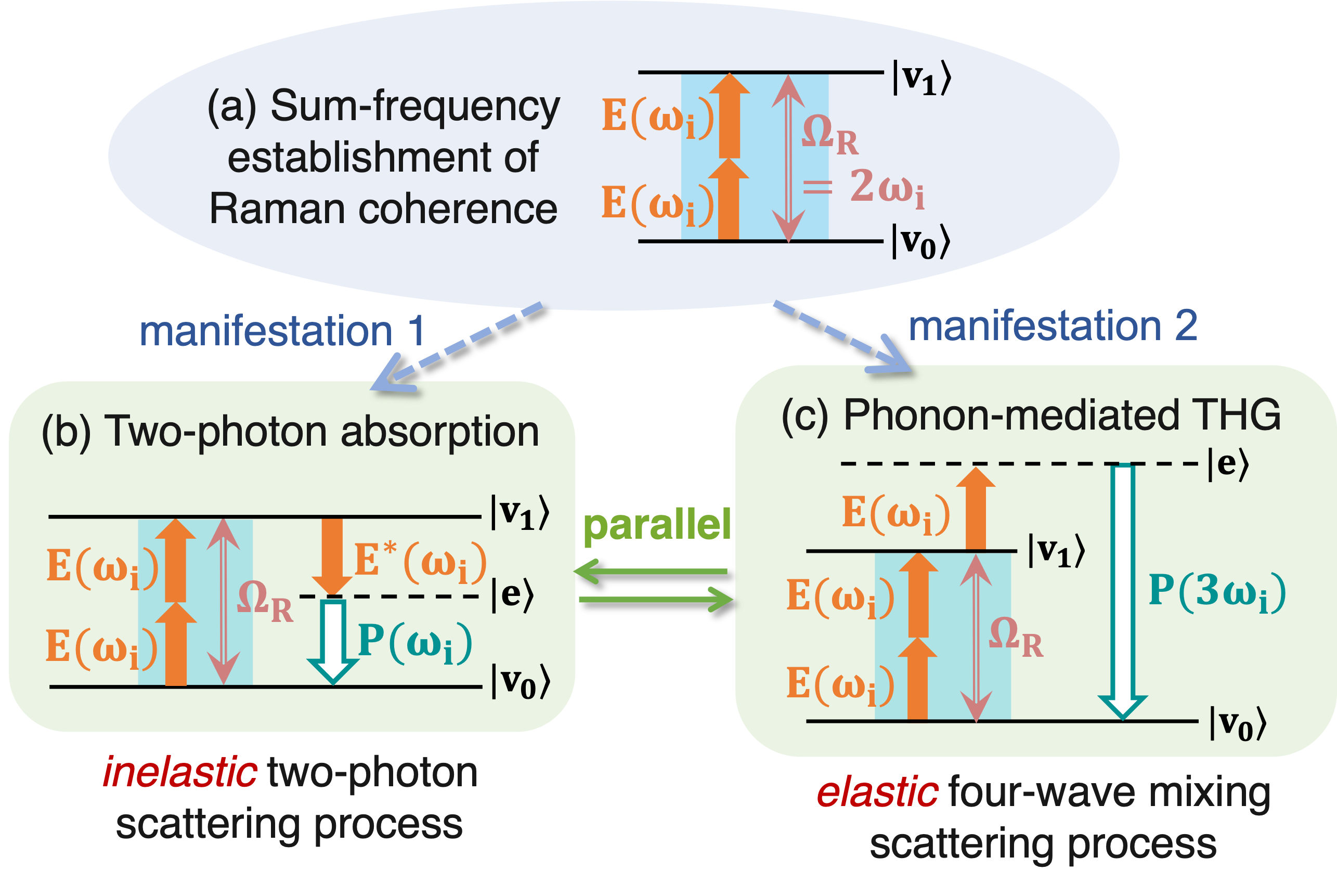}
\vspace{-14 pt}
\caption{\label{fig:level}
(a) Sum-frequency pathway for Raman coherence establishment, as described in \cite{maehrlein2017terahertz}; (b) its associated \textit{inelastic} two-photon scattering process, TPA, and (c) \textit{elastic} four-wave mixing process, PM-THG. $\ket{v_0}$, $\ket{v_1}$ are ground and excited states of a Raman phonon. $\ket{e}$ is a virtual electronic state.
}
\vspace{-18 pt}
\end{figure}

However, when IR-active phonons are absent or Raman-IR coupling is weak, alternative strategies for coherent structural and optical property control are imperative. A Raman phonon coherence can be driven by two light fields 
with either a sum-frequency or difference-frequency beat note resonant with the Raman frequency $\Omega_R$. 
As standard laser frequencies are far higher than phonon frequencies (usually $<$50 THz), the difference-frequency pathway, driven by visible or near-IR light fields, is by far the most familiar case in Raman scattering processes. 
Nonetheless, as strong table-top THz light sources have advanced, a recent development in crystalline solids is to establish Raman coherence via a sum-frequency beat note of the applied THz fields $E(\omega_i)$ -- a frequency range that avoids electronic heating --
where $2\omega_i = \Omega_R$, as Fig.~\ref{fig:level}(a) shows \cite{maehrlein2017terahertz, johnson2019distinguishing,sato2020two}.
Using a femtosecond THz laser, Maehrlein \textit{et al.} \cite{maehrlein2017terahertz} first demonstrated sum-frequency establishment (SFE) of Raman coherence for ultrafast structural modification of diamond.
This pathway was corroborated by Johnson \textit{et al.} \cite{johnson2019distinguishing} in cadmium tungstate via two-dimensional THz spectroscopy, and by Sato \textit{et al.} \cite{sato2020two} in diamond using a free-electron laser. 

Additionally, Refs. \cite{maehrlein2017terahertz,johnson2019distinguishing,sato2020two} discussed two-photon absorption (TPA) of incident THz light (Fig.~\ref{fig:level}(b)) as an inelastic optical scattering effect accompanying SFE of Raman coherence.
This stems from a third-order polarizability at the incident frequency, causing nonlinear absorption of THz pump at a rate proportional to its intensity [\textit{Supplemental Material} (SM), Eq.~7]. 
TPA in this context is analogous to stimulated Raman scattering (SRS), which is an inelastic optical scattering effect that accompanies a \textit{difference-frequency} excitation of a Raman phonon. In the case of SRS, optical scattering is incompletely inelastic, with energy transferred from high frequency to low frequency photons. In the case of TPA, it is completely inelastic. In both cases, the absorbed optical energy promotes an increase in phonon excited-state population. 

In this Letter, we investigate another strong optical effect associated with SFE of Raman coherence, which was not previously observed in solids: a third-harmonic generation (THG) process enhanced by the intermediate Raman coherence, which we refer to as phonon-mediated THG (PM-THG) (Fig.~\ref{fig:level}(c)). Just as TPA relates to SRS, PM-THG may be regarded as a low-frequency analog of coherent anti-Stokes Raman scatterering (CARS), with Raman coherence established by a sum-frequency pathway rather than the conventional difference-frequency pathway. When the applied fields are degenerate, the `anti-Stokes' frequency, $\omega_a = \omega_i+\Omega_R$, is equivalent to the third harmonic of the incident laser frequency, $3\omega_i$. Like CARS, PM-THG is an elastic four-wave mixing process where the Raman coherence increases the optical scattering cross-section but \emph{all incident energy is returned to the optical fields}. Note that TPA and PM-THG, as different manifestations of SFE of Raman coherence, are parallel processes that occur simultaneously. (See Sec.~\uppercase\expandafter{\romannumeral1} in SM for further comparison and analysis of the four optical scattering effects discussed above.) 

Leveraging the sufficiently narrow bandwidth and high peak intensity of a home-built picosecond THz source, we present the first experimental demonstration and theoretical analysis of PM-THG in solid-state systems, showcasing attributes distinct from those previously observed in molecular systems \cite{she1975infrared,kildal1976infrared,kildal1977resonant,miyamoto2017vibrational,kinder2021detection}. 
Following Refs.~\cite{maehrlein2017terahertz,sato2020two}, we study diamond, an important photonic material \cite{aharonovich2011diamond,hausmann2014diamond,janitz2020cavity, abulikemu2021second}
without IR-active phonons. 
As PM-THG in diamond is highly sensitive to field polarization, the contributions of the purely electronic and phonon-mediated pathways to THG near resonance can be disentangled. This allows detection of a remarkable THG enhancement by over 100-fold at resonance that can be further extended to 3000-fold. 
Moreover, a study of frequency dependence uncovers a new phenomenon, the suppression of THG above resonance due to out-of-phase purely electronic and phonon-mediated contributions to the polarization field, allowing a six-order-of-magnitude tuning range of the THG efficiency.
These observations thus illuminate new opportunities for strongly linked structural and optical property modification of crystals with light at low (THz/mid-infrared) frequencies, having relevance both to photonics applications as a strongly enhanced or suppressed solid-state optical wave-mixing nonlinearity, and to condensed-phase physics as a new method for detecting structural dynamics during light-driven material phase control and for spectroscopy in the THz regime.

Our experiment employed an 80-$\mu$m thick type-IIa high-purity diamond crystal plate with [001] orientation grown by microwave plasma chemical vapor deposition (Applied Diamond, Inc.) 
We focus on the sole Raman-active optical phonon of diamond with F$_{2g}$ symmetry, as investigated in \cite{ishioka2006coherent, maehrlein2017terahertz,sato2020two}. 
We determined the phonon resonant frequency ($f_R = \Omega_R/2\pi = 39.952\pm 0.002$ THz) and linewidth ($\Gamma/2\pi = 0.056\pm0.003$ THz) using a 
confocal Raman microscope, and found them to be consistent with previous reports \cite{maehrlein2017terahertz,solin1970raman,ishioka2006coherent,levenson1972interference} (SM, Sec.~\uppercase\expandafter{\romannumeral2}). To maximize coherent growth of the third-harmonic field, the 80-$\mu$m sample thickness was chosen to match the THG coherence length, $L_{coh}=|\pi c/ \left[3 \omega_i (n(\omega_i) - n(3\omega_i))\right]|$, where $c$ is the speed of light and $n$ is the refractive index.

PM-THG requires an applied laser frequency $f_i$ at half the phonon frequency, $\sim$20 THz. 
We used a home-built tabletop tunable THz source with $<$0.5 THz full-width at half-maximum  (FWHM) bandwidth, which is narrower than the pulses used in \cite{maehrlein2017terahertz, sato2020two}. This bandwidth ensures that a large fraction of the laser power is directed towards Raman phonon excitation, which is essential for observing strong PM-THG.
This THz source combines adiabatic difference frequency generation \cite{suchowski2013octave,krogen2017generation} with programmable pulse-shaping to generate THz pulses with customizable bandwidth and central frequency continuously tunable from 14 to 37 THz \cite{chang2021flexible}.

\begin{figure}
\centering
\includegraphics[width=8.5 cm]
{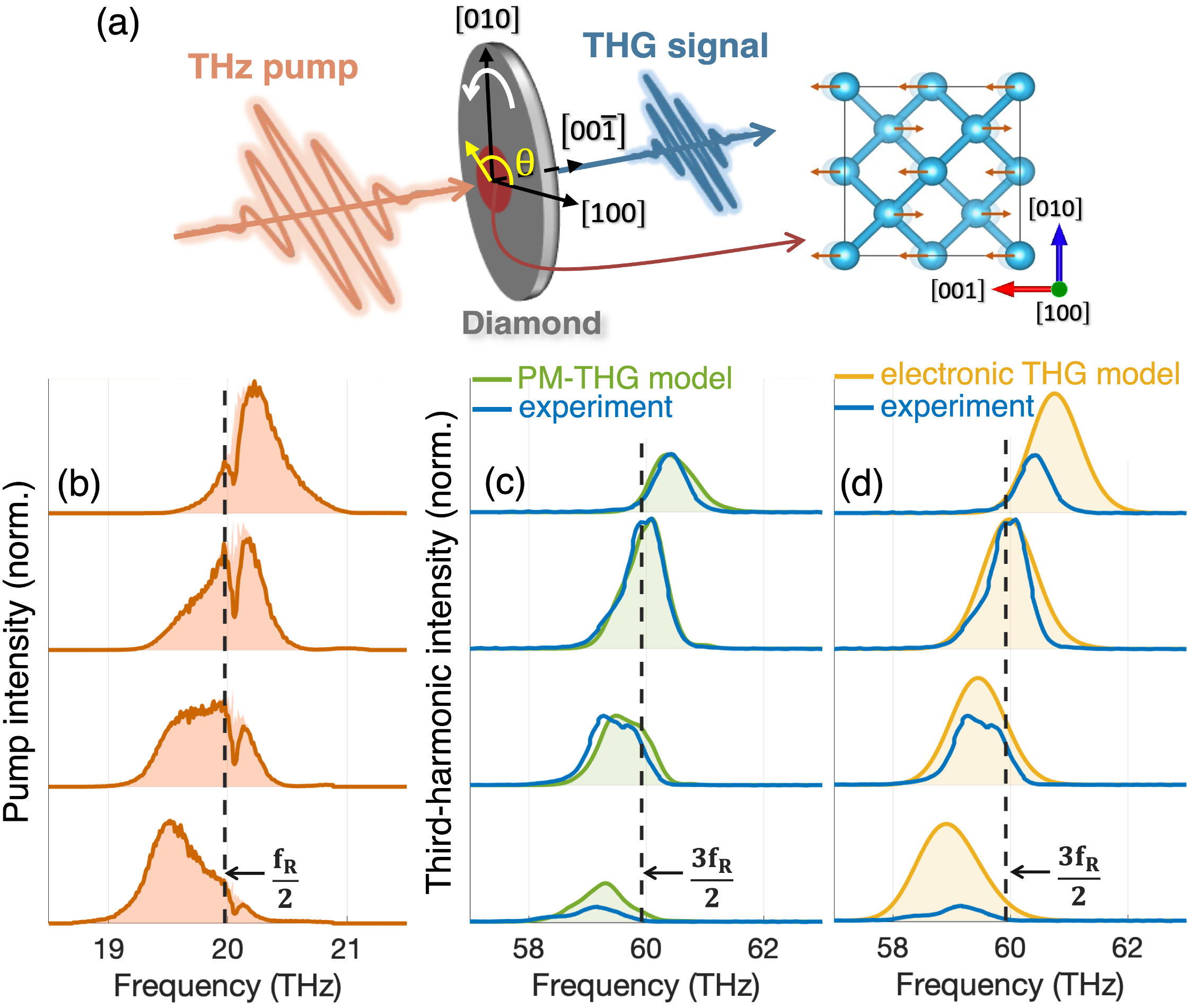}
\vspace{-10 pt}
\caption{(a) Orientations of field polarizations, diamond crystal axes, and F$_{2g}$ phonon of diamond. Orange arrows depict motion of carbon atoms (blue circles) in a phonon oscillation. Measured spectral intensities of (b) THz pump and (c),(d) THG signals (blue), compared to theoretical predictions from the PM-THG model (green) and electronic THG model (yellow), uniformly normalized across rows such that theoretical and experimental maxima match in the second row.
}
\vspace{-16 pt}
\label{fig:frequency detuning}
\end{figure}

A 15-mm focal length \ce{ZnSe} lens focused the THz pump onto the diamond sample, resulting in a maximum fluence of 13 \si{mJ/cm^2} ($\sim$1-ps pulse duration and up to 1.5-\si{\mu J} pulse energy).  Rotation of the diamond plate allowed the electric field of the linearly polarized pump to align parallel to any angle $\theta$ in the [100]-[010] crystal plane (Fig.~\ref{fig:frequency detuning}(a)).  
Emitted light was collimated by a \ce{ZnSe} lens and passed through two \ce{CaF2} windows to block the THz pump before passing into a \ce{HgCdTe} spectrometer. 
The THz generation stage and the path of the THz beam until reaching the diamond plate were purged by \ce{N2} gas to reduce absorption from \ce{CO2} in air.

We observed a sharp increase in emitted light at the third-harmonic of the incident frequency, $3f_i$, and its strong sensitivity to the pump frequency and polarization angle in the vicinity of the resonant condition ($f_i = f_R/2 =$ 19.98 THz). Fig.~\ref{fig:frequency detuning}(b)-(d) shows the incident pump and emitted third-harmonic spectra, obtained with the crystal plate oriented to maximize the observed THG power. The pump central frequencies were adjusted within a range close to resonance of 19.6--20.4 THz. 
To corroborate the phonon-mediated origin of the observed THG spectra, 
we compare them to models based on solely phonon-mediated or purely electronic pathways. For low-frequency fields in diamond, the third-order electric polarization has two components at lowest order,
\begin{eqnarray}
 \nonumber
P^{(3)}(t) &=&  \chi_e^{(3)} E(t)E(t)E(t) + \Pi Q_{R}(t) E(t) \\
&\equiv& P_e^{(3)}(t)+P_R^{(3)}(t).
\label{eq:P3_total}
\end{eqnarray}
The first term $P_e^{(3)}(t)$ is the origin of non-resonant electronic THG, where $\chi_e^{(3)}$ is the approximately instantaneous third-order electronic susceptibility arising from the anharmonic electronic potential. The second term, $P_R^{(3)}(t)$, is the product of the incident electric field $E(t)$ and the Raman phonon displacement $Q_R(t)$, describing the origin of PM-THG. The Raman phonon dynamics can be modeled as a classical Lorentzian oscillator with equation of motion 
\begin{equation}
\begin{aligned}
M\left(\ddot{Q}_R + 2 \Gamma \dot{Q}_R+\Omega_R^{2}  Q_R\right)
= \Pi_{\alpha\beta} E_\alpha E_\beta,
\label{eq:eom}
\end{aligned}
\end{equation}
where $M$ is the phonon effective mass. 
The phonon-to-electric-field coupling originates from the dependence of the linear electronic susceptibility $\chi_e^{(1)}$ on $Q_R$, given by the Raman polarizability tensor $\Pi$. For pump field propagating along the diamond [001] axis and polarized at an angle $\theta$, such that $\vec E(t) = E(t) \left[\begin{matrix}\cos \theta  & \sin \theta & 0\end{matrix}\right]$ with $\theta=$ 0° corresponding to the [100] direction, this second-rank tensor takes the form \cite{solin1970raman}, $\Pi = \left.\frac{\partial \chi_e^{(1)}}{\partial Q_R}\right|_{Q_R=0} = \Pi_{0}\big(\begin{smallmatrix}
0 & 1 & 0\\
1 & 0 & 0\\
0 & 0 & 0
\end{smallmatrix}\big)$. 

We solve Eqs.~\ref{eq:P3_total},\ref{eq:eom} in the frequency domain and derive the Fourier transforms of $P_e^{(3)}(t)$ and $P_R^{(3)}(t)$ when $E(t)$ is polarized along the [110] axis (SM, Sec.~\uppercase\expandafter{\romannumeral4}):  
\begin{equation}
\tilde P_e^{(3)}(\omega) = \left[\chi_e^{(3)}\left(\tilde E(\omega) * \tilde E(\omega)\right)\right]* \tilde E(\omega),
\label{eq:P_e_omega}
\end{equation}
\begin{equation}
\begin{aligned}
\tilde P_R^{(3)}(\omega) &= \Pi_0 \tilde Q_{R}(\omega) * \tilde E(\omega)  \\
&= \left[\chi_R^{(3)}(\omega)\left(\tilde E(\omega) * \tilde E(\omega)\right)\right]* \tilde E(\omega).
\label{eq:P_R_omega}
\end{aligned}
\end{equation}
$\tilde E(\omega)$, $\tilde Q_{R}(\omega)$ are the Fourier transforms of $E(t)$, $Q(t)$, and $*$ denotes linear convolution. We defined a Raman susceptibility, $\chi_R^{(3)}(\omega) = \Pi_0^2/M \left(\Omega_R^2-\omega^2-2i \Gamma \omega \right)$.

\begin{figure}
\includegraphics[width=8.5cm]
{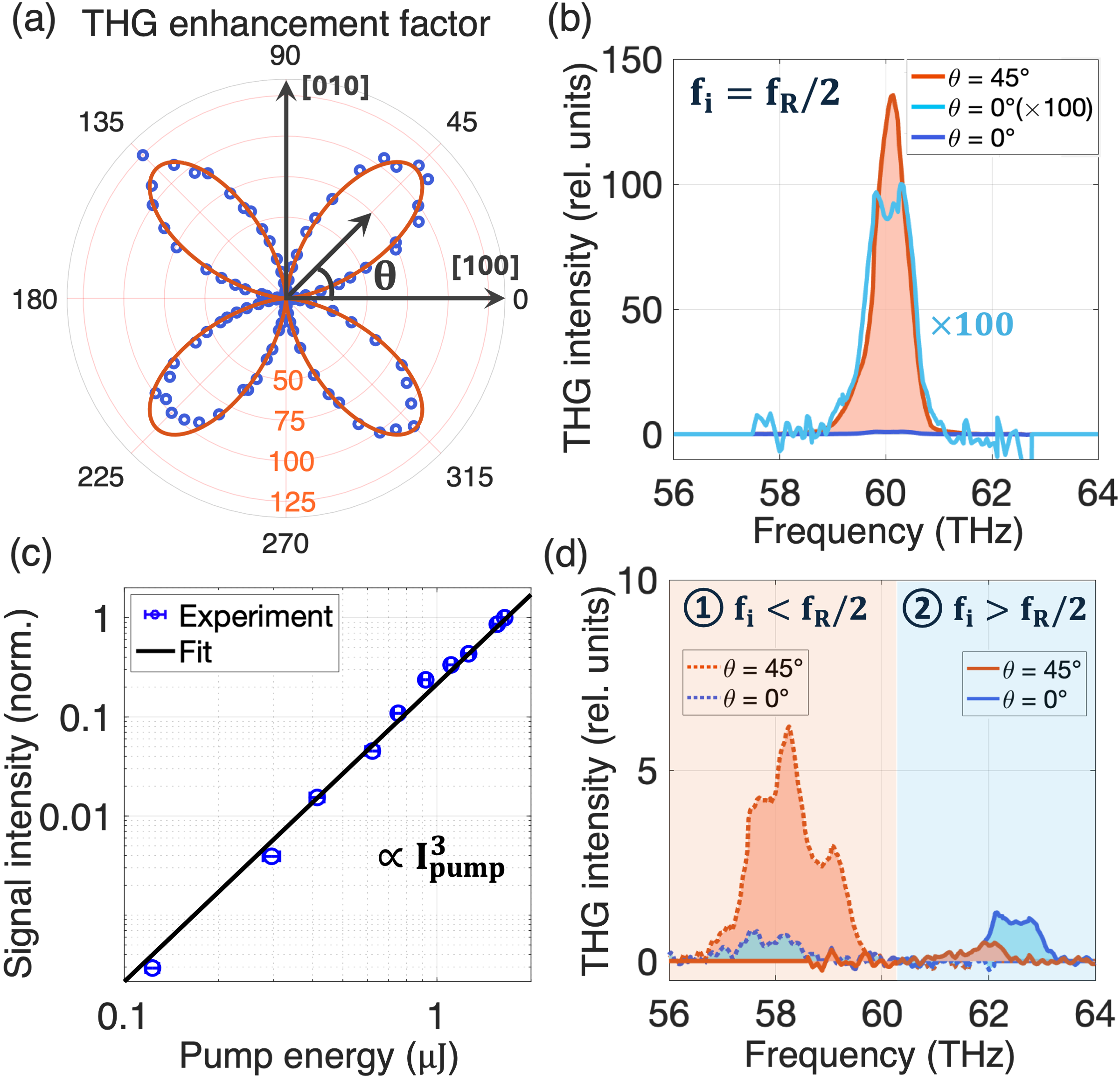}
\vspace{-10 pt}
\caption{\label{fig:angle}(a) Polar plot of the THG power enhancement factor 
(blue circles) vs. pump polarization angle $\theta$ with pump centered at resonance, with best fit to $113\sin^2(2\theta)$ (red curve). 
(b) THG signal spectra observed at $\theta=$ 45° (red) and 0° (dark blue for actual intensity, light blue for intensity magnified by 100). 
(c) Signal intensity dependence on pump energy at $\theta =$ 45° (log-log scale), with a black line of slope 3 for reference. (d) THG signal spectra observed at $\theta=$ 45° and 0° for pump frequencies below (dotted lines) and above (solid lines) resonance,
with corresponding pump spectra shown in Fig.~S4 (SM). All the signal spectra in (b) and (d) are normalized to the maximum signal intensity at $\theta =$ 0° in (b).
}
\vspace{-16 pt}
\end{figure}

To compare the measurements against theory, we incorporated the pump spectra (Fig.~\ref{fig:frequency detuning}(b)) into our theoretical model, and calculated signal spectra corresponding to the PM-THG (Fig.~\ref{fig:frequency detuning}(c), green) and electronic-THG (Fig.~\ref{fig:frequency detuning}(d), yellow) mechanisms acting independently. 
[We note, a consistent dip at 20.1 THz in pump spectra resulted from \ce{CO2} absorption along the spectrometer measurement path that was impractical to purge with \ce{N2}. To remove this artefact, a Lorentzian absorption model was used to estimate the true pump spectrum at the diamond crystal (Fig.~\ref{fig:frequency detuning}(b),  shaded region).]
While the PM-THG model reproduces both the experimental trend in signal intensity and spectral shape, the electronic-THG model predicts a signal intensity lacking sensitivity to pump frequency and a broader signal spectrum centered further from resonance than the measurement. 
The differences between the theoretical predictions can be understood through the form of the nonlinear polarization (Eq.~\ref{eq:P_e_omega} vs.  Eq.~\ref{eq:P_R_omega}), whereby PM-THG contains an implicit convolution between the frequency-dependent Raman coherence $\tilde Q_R(\omega)$ and $\tilde E(\omega)$, but purely electronic THG involves only a double autoconvolution of $\tilde E(\omega)$.

Next, by investigating the dependence of the THG power (integrated power spectrum) on pump polarization angle $\theta$, we can both distinguish the relative contributions of the purely electronic and phonon-mediated pathways and reveal interference effects in the total third-order polarization. Fig.~\ref{fig:angle}(a) summarizes THG power dependence on pump polarization with the pump frequency $f_i$ centered at the $f_R/2$ resonance.
The radial coordinate represents the ratio of third-harmonic signal power to the average of its minima found at $\theta=$ 0°, 90°, 180°, and 270°. Strong maxima, exceeding 100, are observed at $\theta =$ 45°, 135°, 225° and 315°. Fig.~\ref{fig:angle}(b) compares the signal spectra at $\theta =$ 45° and $\theta =$ 0°. Following the theoretical discussion above, the amplitude of Raman oscillation $Q_R \propto \sin(2\theta)$. As the experimental data in Fig.~\ref{fig:angle}(a) is observed to fit closely to a $\sin^2 (2\theta)$ function (red curve), we can thus infer that the signal intensity is dominated by the phonon-mediated contribution to the third-order polarizability, $|P^{(3)}|^2 \propto |Q_R|^2 $.
The non-resonant electronic contribution, in contrast, exhibits near $\theta$-independence, as determined by the $\chi^{(3)}_e$ tensor of diamond crystal \cite{levenson1974dispersion}
(SM, Sec.~\uppercase\expandafter{\romannumeral3}). Thus, the minimal signal observed at $\theta = 0$° and 90° originates only from the purely electronic pathway, allowing interpretation of the data in Fig.~\ref{fig:angle}(a) as a THG `enhancement factor' relative to the purely electronic effect. A best fit of $113\sin^2 (2\theta)$ to this data indicates that the phonon-mediated pathway enhances THG intensity by at least a factor of 113.

Fig.~\ref{fig:angle}(c) shows the dependence of the third-harmonic signal intensity on incident pump energy at $\theta = 45$°. On a log-log scale, the data fits well to a line of slope 3, confirming a THG process as the signal origin and indicating a largely-undiminished THz pump and unsaturated phonon oscillation amplitude.  
This allows interpretation of the measured THG enhancement factor as $|(P_R^{(3)}+P_e^{(3)})/P_e^{(3)}|^2$, from which we deduce a susceptibility ratio at resonance $| \chi_R^{(3)}(\Omega_R)/\chi_e^{(3)}| \gtrsim 58$ (SM, Sec.~\uppercase\expandafter{\romannumeral4}). 

Fig.~\ref{fig:angle}(a,b) demonstrates the dominance of the PM-THG pathway with a resonant pump. However, 
detuning the pump frequency reveals a more complex THG phenomenon. 
As Fig.~\ref{fig:angle}(d) shows, for a pump with central frequency $f_i<f_R/2$, the THG intensity observed at $\theta = 45$° exceeds that at $\theta=0$° by only a factor of 10, a magnitude smaller than the enhancement observed at resonance.
Remarkably, for pump central frequency $f_i>f_R/2$, the THG intensity at $\theta = 45$° is suppressed below that at $\theta=0$°. This indicates destructive interference between the phonon-mediated and electronic pathways, resulting in a total THG intensity lower than that from the purely electronic contribution alone.

\begin{figure}
\includegraphics[width=8.5 cm]
{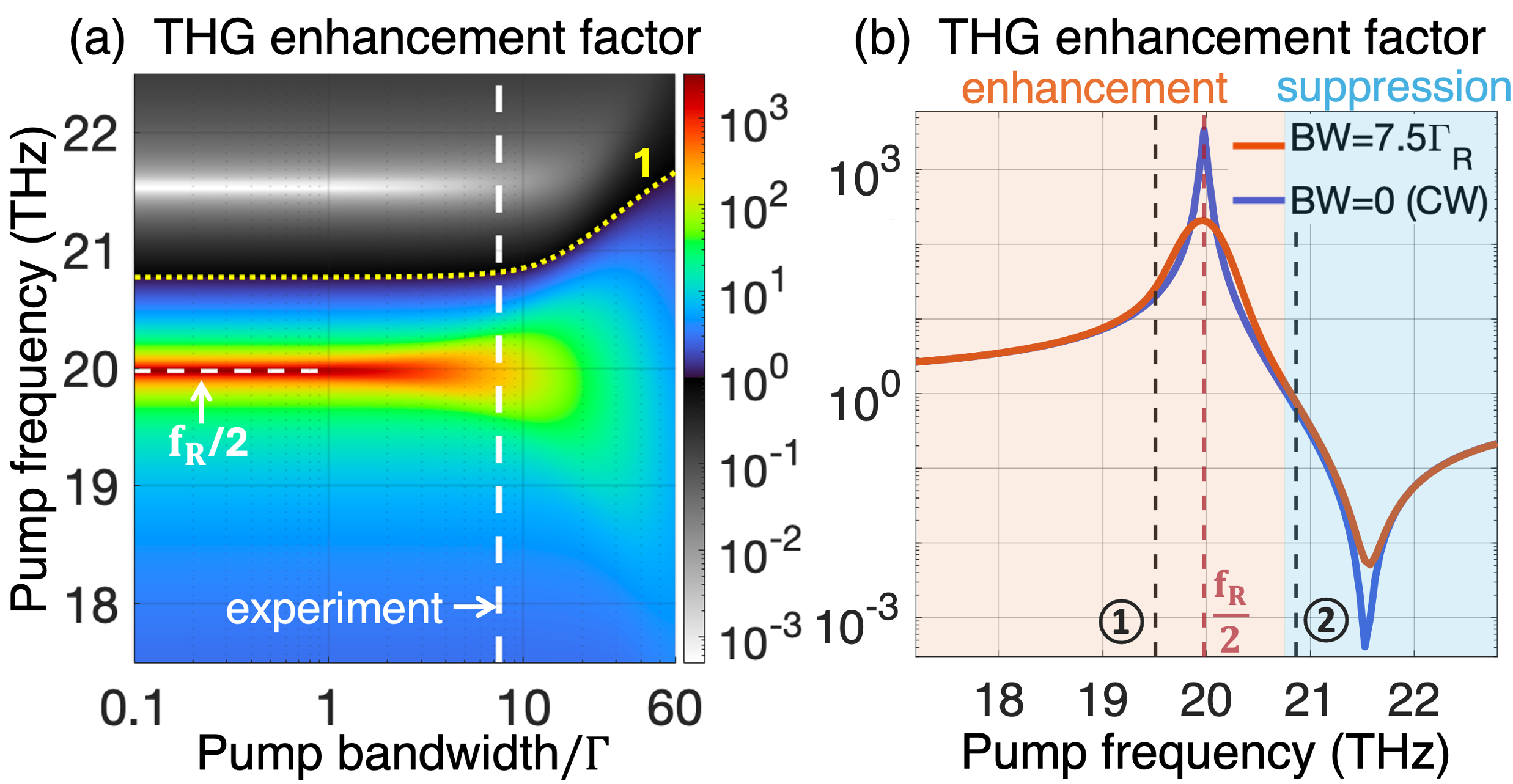}
\vspace{-8 pt}
\caption{\label{fig:extrapolation} (a) THG enhancement factor as a function of pump central frequency and bandwidth, spanning from $>$10$^3$ to $<$10$^{-3}$.
The yellow dotted contour line (enhancement factor = 1) divides the figure into two regions: THG enhancement (below) and suppression (above).
(b) THG enhancement factor vs. pump central frequency for a continuous-wave (CW) pump (blue) and a Gaussian-shape pump with a bandwidth of $7.5 \Gamma$ (red), simulating our experimental conditions. Red and blue shading indicates THG enhancement and suppression regions respectively. Black dashed lines mark pump central frequencies corresponding to the signal \ding{172} and \ding{173} in Fig.~\ref{fig:angle}(d).
}
\vspace{-16 pt}
\end{figure}

To explore this possibility, we calculate the THG enhancement factor, $|(P_R^{(3)}+P_e^{(3)})/P_e^{(3)}|^2$, using Eqs.~\ref{eq:P_e_omega},\ref{eq:P_R_omega} and the obtained susceptibility ratio $| \chi_R^{(3)}(\Omega_R)/\chi_e^{(3)}| = 58$. Fig.~\ref{fig:extrapolation}(a) plots our results against pump central frequency and FWHM bandwidth. 
We observe several notable features. First, on resonance ($f_i=f_R/2$), a much larger enhancement is possible than in our experiment. 
Our experimental pump bandwidth (marked by the vertical dashed line) was $\sim$7-8 times wider than the Raman mode linewidth $\Gamma$, resulting in an enhancement factor $>100$ on resonance. However, for pump bandwidths narrower than $\Gamma$ (corresponding to pulse durations $>$10 ps), the enhancement factor approaches its maximum of $| \chi_R^{(3)}(\Omega_R)/\chi_e^{(3)}|^2 \simeq 3300$, demonstrating the great potential of lattice vibrations to modify the optical properties of crystals. 
Conversely, for a broader pump bandwidth of $60\Gamma$ (corresponding to 100-fs pulse duration, as used in \cite{maehrlein2017terahertz}), the predicted enhancement factor falls to $\sim$5.

Second, while the THG enhancement factor drops rapidly as the pump central frequency deviates from resonance ($f_R/2$) as expected,  this decline is faster with frequency detuning above $f_R/2$ than below it,
eventually reaching a region with enhancement factor $<1$. A yellow dotted contour line, denoting a factor of 1,  
thus divides Fig.~\ref{fig:extrapolation}(a) into two regions: THG enhancement (below) and suppression (above). 
This trend is further illustrated by Fig.~\ref{fig:extrapolation}(b), showing the frequency-dependent enhancement factor for a continuous-wave pump and a pump with a bandwidth of $7.5 \Gamma$, which simulates our experimental conditions. 
Suppression occurs when the pump central frequency is detuned beyond 20.8 THz. The pump frequencies corresponding to the signal shown in Fig.~\ref{fig:angle}(d), marked by lines \ding{172} and \ding{173}, lie within the regions of enhancement (red shading) and suppression (blue shading) respectively. 
For a pump bandwidth significantly narrower than the Raman linewidth, the total THG intensity can be suppressed to below $10^{-3}$ times that of the purely electronic pathway at a pump frequency of $\sim$21.5 THz.
This nearly full cancellation of the nonlinear polarization fields contributed by the phonon-mediated and electronic pathways is explained by the opposite signs of $\chi_e^{(3)}$ and the real part of $\chi_R^{(3)}(\omega)$ above resonance (SM, Sec.~\uppercase\expandafter{\romannumeral5}). 
Similar interference behaviour has been observed between purely electronic and CARS pathways for four-wave mixing between visible frequency waves in diamond \cite{levenson1972interference}. Our first demonstration of this phenomenon in THG thus illustrates the correspondence between optical scattering physics in the THz domain via SFE of Raman coherence and those via traditional difference-frequency excitation with high-frequency light. 



In summary, we demonstrate strong anisotropic third-harmonic generation in diamond mediated by resonant sum-frequency driving of Raman phonons with THz light. The observed pump frequency and polarization dependence of the signal confirms the dominant role of PM-THG over purely electronic THG at half the Raman resonance. This polarization dependence -- unique to condensed systems -- allows the phonon-mediated and purely electronic contributions to be resolved. At resonance, we observed over 100-fold enhancement in THG efficiency, and predict over 3000-fold enhancement for narrower-bandwidth pumping. 
This corresponds to a nonlinear susceptibility of PM-THG surpassing that of purely electronic THG by over 58 times. Moreover, we have discovered THG suppression above resonance, a result of destructive interference between phonon-mediated and purely electronic contributions to the polarization field. 
These findings highlight the intimate tie between the nonlinear optical response and the structural response of diamond, with timely relevance to several fields of research.

From a photonics perspective, diamond has attracted considerable recent interest as a next-generation photonic material due to excellent linear and nonlinear optical properties and relevance to quantum information science \cite{aharonovich2011diamond,hausmann2014diamond,janitz2020cavity, abulikemu2021second}.  
In virtue of its strongly enhanced frequency upconversion susceptibility, PM-THG adds additional capability to the diamond photonic platform. The phonon-mediated pathway may also allow higher-order (e.g., $5^{th}, 7^{th}, 9^{th}$) harmonic generation processes to become efficient, potentially bridging the incident field near 20 THz to the telecom range. Requiring only a Raman phonon, PM-THG is of potential use to any condensed-phase photonics platform. 
Moreover, the strong dependence of the THG efficiency on pump polarization direction and frequency (up to $>$6 orders of magnitude) suggests new applications in areas such as infrared nonlinear optical switching and orientation diagnostics. As a degenerate low-frequency analog of CARS, PM-THG can also be used for spectroscopic applications amenable to THz and mid-infrared light, with the unprecedented simplicity of requiring only a single incident frequency compared to stimulated Raman spectroscopies,  
while sharing CARS' advantage of producing a signal at a frequency distinct from the incident light, thus easing detection. Beyond PM-THG, non-degenerate instances of phonon-mediated four-wave mixing could be employed.  
For example, applying two THz frequencies with a sum-frequency beat note slightly detuned from $\Omega_R$ might induce significant cross-phase modulation, markedly affecting the refractive indices of the incident fields.

Of significance to the light-driven structural control of quantum materials, 
our report clarifies that PM-THG and TPA are parallel optical polarization effects -- elastic and inelastic, respectively -- that both accompany THz sum-frequency driving of Raman phonons in crystals. 
Our findings demonstrate the potential of Raman phonons to induce marked changes in the optical properties of solids, 
complementing recent investigations on the changes tied to IR-active phonon nonlinearities \cite{dolgaleva2015prediction, khalsa2021ultrafast, ginsberg2023phonon, zibod2023strong}. Moreover, the duality of structural and optical responses in materials allows PM-THG to serve as a convenient marker of concurrent structural changes during light-driven phase transitions.
In future work, in analogy to time-domain anti-Stokes Raman scattering \cite{sato2020two}, the temporal evolution of the coherent phonon amplitude might be resolved by applying a pump and time-delayed probe at the same frequency of $f_R/2$ with a small angular separation, and detecting the resulting noncollinear PM-THG signal. This technique eliminates the requirement for carrier-envelope-phase stable THz pump pulses. 
Thus, beyond its value as a strong nonlinear optical response, PM-THG offers a powerful laser-lab-scale diagnostic for the growing field of light-driven structural control in condensed matter physics.

\begin{acknowledgments}
This research was primarily supported through the Cornell University Materials Research Science and Engineering Center DMR-1719875. Support for instrumentation was provided by the Kavli Institute at Cornell.
\end{acknowledgments}

\bibliographystyle{apsrev}
\bibliography{apssamp}

\begin{thebibliography}{24}
\expandafter\ifx\csname natexlab\endcsname\relax\def\natexlab#1{#1}\fi
\expandafter\ifx\csname bibnamefont\endcsname\relax
  \def\bibnamefont#1{#1}\fi
\expandafter\ifx\csname bibfnamefont\endcsname\relax
  \def\bibfnamefont#1{#1}\fi
\expandafter\ifx\csname citenamefont\endcsname\relax
  \def\citenamefont#1{#1}\fi
\expandafter\ifx\csname url\endcsname\relax
  \def\url#1{\texttt{#1}}\fi
\expandafter\ifx\csname urlprefix\endcsname\relax\def\urlprefix{URL }\fi
\providecommand{\bibinfo}[2]{#2}
\providecommand{\eprint}[2][]{\url{#2}}

\bibitem[{\citenamefont{Disa et~al.}(2021)\citenamefont{Disa, Nova, and Cavalleri}}]{disa2021engineering}
\bibinfo{author}{\bibfnamefont{A.~S.} \bibnamefont{Disa}}, \bibinfo{author}{\bibfnamefont{T.~F.} \bibnamefont{Nova}}, \bibnamefont{and} \bibinfo{author}{\bibfnamefont{A.}~\bibnamefont{Cavalleri}}, \bibinfo{journal}{Nat. Phys.} \textbf{\bibinfo{volume}{17}}, \bibinfo{pages}{1087} (\bibinfo{year}{2021}).

\bibitem[{\citenamefont{Dolgaleva et~al.}(2015)\citenamefont{Dolgaleva, Materikina, Boyd, and Kozlov}}]{dolgaleva2015prediction}
\bibinfo{author}{\bibfnamefont{K.}~\bibnamefont{Dolgaleva}}, \bibinfo{author}{\bibfnamefont{D.~V.} \bibnamefont{Materikina}}, \bibinfo{author}{\bibfnamefont{R.~W.} \bibnamefont{Boyd}}, \bibnamefont{and} \bibinfo{author}{\bibfnamefont{S.~A.} \bibnamefont{Kozlov}}, \bibinfo{journal}{Phys. Rev. A} \textbf{\bibinfo{volume}{92}}, \bibinfo{pages}{023809} (\bibinfo{year}{2015}).

\bibitem[{\citenamefont{Khalsa et~al.}(2021)\citenamefont{Khalsa, Benedek, and Moses}}]{khalsa2021ultrafast}
\bibinfo{author}{\bibfnamefont{G.}~\bibnamefont{Khalsa}}, \bibinfo{author}{\bibfnamefont{N.~A.} \bibnamefont{Benedek}}, \bibnamefont{and} \bibinfo{author}{\bibfnamefont{J.}~\bibnamefont{Moses}}, \bibinfo{journal}{Phys. Rev. X} \textbf{\bibinfo{volume}{11}}, \bibinfo{pages}{021067} (\bibinfo{year}{2021}).

\bibitem[{\citenamefont{Ginsberg et~al.}(2023)\citenamefont{Ginsberg, Jadidi, Zhang, Chen, Tancogne-Dejean, Chae, Patwardhan, Xian, Watanabe, Taniguchi et~al.}}]{ginsberg2023phonon}
\bibinfo{author}{\bibfnamefont{J.~S.} \bibnamefont{Ginsberg}}, \bibinfo{author}{\bibfnamefont{M.~M.} \bibnamefont{Jadidi}}, \bibinfo{author}{\bibfnamefont{J.}~\bibnamefont{Zhang}}, \bibinfo{author}{\bibfnamefont{C.~Y.} \bibnamefont{Chen}}, \bibinfo{author}{\bibfnamefont{N.}~\bibnamefont{Tancogne-Dejean}}, \bibinfo{author}{\bibfnamefont{S.~H.} \bibnamefont{Chae}}, \bibinfo{author}{\bibfnamefont{G.~N.} \bibnamefont{Patwardhan}}, \bibinfo{author}{\bibfnamefont{L.}~\bibnamefont{Xian}}, \bibinfo{author}{\bibfnamefont{K.}~\bibnamefont{Watanabe}}, \bibinfo{author}{\bibfnamefont{T.}~\bibnamefont{Taniguchi}}, \bibnamefont{et~al.}, \bibinfo{journal}{Nat. Commun.} \textbf{\bibinfo{volume}{14}}, \bibinfo{pages}{7685} (\bibinfo{year}{2023}).

\bibitem[{\citenamefont{Zibod et~al.}(2023)\citenamefont{Zibod, Rasekh, Yildrim, Cui, Bhardwaj, M{\'e}nard, Boyd, and Dolgaleva}}]{zibod2023strong}
\bibinfo{author}{\bibfnamefont{S.}~\bibnamefont{Zibod}}, \bibinfo{author}{\bibfnamefont{P.}~\bibnamefont{Rasekh}}, \bibinfo{author}{\bibfnamefont{M.}~\bibnamefont{Yildrim}}, \bibinfo{author}{\bibfnamefont{W.}~\bibnamefont{Cui}}, \bibinfo{author}{\bibfnamefont{R.}~\bibnamefont{Bhardwaj}}, \bibinfo{author}{\bibfnamefont{J.-M.} \bibnamefont{M{\'e}nard}}, \bibinfo{author}{\bibfnamefont{R.~W.} \bibnamefont{Boyd}}, \bibnamefont{and} \bibinfo{author}{\bibfnamefont{K.}~\bibnamefont{Dolgaleva}}, \bibinfo{journal}{Adv. Opt. Mater.} \textbf{\bibinfo{volume}{11}}, \bibinfo{pages}{2202343} (\bibinfo{year}{2023}).

\bibitem[{\citenamefont{Maehrlein et~al.}(2017)\citenamefont{Maehrlein, Paarmann, Wolf, and Kampfrath}}]{maehrlein2017terahertz}
\bibinfo{author}{\bibfnamefont{S.}~\bibnamefont{Maehrlein}}, \bibinfo{author}{\bibfnamefont{A.}~\bibnamefont{Paarmann}}, \bibinfo{author}{\bibfnamefont{M.}~\bibnamefont{Wolf}}, \bibnamefont{and} \bibinfo{author}{\bibfnamefont{T.}~\bibnamefont{Kampfrath}}, \bibinfo{journal}{Phys. Rev. Lett.} \textbf{\bibinfo{volume}{119}}, \bibinfo{pages}{127402} (\bibinfo{year}{2017}).

\bibitem[{\citenamefont{Johnson et~al.}(2019)\citenamefont{Johnson, Knighton, and Johnson}}]{johnson2019distinguishing}
\bibinfo{author}{\bibfnamefont{C.~L.} \bibnamefont{Johnson}}, \bibinfo{author}{\bibfnamefont{B.~E.} \bibnamefont{Knighton}}, \bibnamefont{and} \bibinfo{author}{\bibfnamefont{J.~A.} \bibnamefont{Johnson}}, \bibinfo{journal}{Phys. Rev. Lett.} \textbf{\bibinfo{volume}{122}}, \bibinfo{pages}{073901} (\bibinfo{year}{2019}).

\bibitem[{\citenamefont{Sato et~al.}(2020)\citenamefont{Sato, Yoshida, Zen, Hachiya, Goto, Sagawa, and Ohgaki}}]{sato2020two}
\bibinfo{author}{\bibfnamefont{O.}~\bibnamefont{Sato}}, \bibinfo{author}{\bibfnamefont{K.}~\bibnamefont{Yoshida}}, \bibinfo{author}{\bibfnamefont{H.}~\bibnamefont{Zen}}, \bibinfo{author}{\bibfnamefont{K.}~\bibnamefont{Hachiya}}, \bibinfo{author}{\bibfnamefont{T.}~\bibnamefont{Goto}}, \bibinfo{author}{\bibfnamefont{T.}~\bibnamefont{Sagawa}}, \bibnamefont{and} \bibinfo{author}{\bibfnamefont{H.}~\bibnamefont{Ohgaki}}, \bibinfo{journal}{Phys. Lett. A} \textbf{\bibinfo{volume}{384}}, \bibinfo{pages}{126223} (\bibinfo{year}{2020}).

\bibitem[{\citenamefont{She and Billman}(1975)}]{she1975infrared}
\bibinfo{author}{\bibfnamefont{C.}~\bibnamefont{She}} \bibnamefont{and} \bibinfo{author}{\bibfnamefont{K.~W.} \bibnamefont{Billman}}, \bibinfo{journal}{Appl. Phys. Lett.} \textbf{\bibinfo{volume}{27}}, \bibinfo{pages}{76} (\bibinfo{year}{1975}).

\bibitem[{\citenamefont{Kildal and Deutsch}(1976)}]{kildal1976infrared}
\bibinfo{author}{\bibfnamefont{H.}~\bibnamefont{Kildal}} \bibnamefont{and} \bibinfo{author}{\bibfnamefont{T.}~\bibnamefont{Deutsch}}, \bibinfo{journal}{IEEE J. Quantum Electron.} \textbf{\bibinfo{volume}{12}}, \bibinfo{pages}{429} (\bibinfo{year}{1976}).

\bibitem[{\citenamefont{Kildal and Brueck}(1977)}]{kildal1977resonant}
\bibinfo{author}{\bibfnamefont{H.}~\bibnamefont{Kildal}} \bibnamefont{and} \bibinfo{author}{\bibfnamefont{S.}~\bibnamefont{Brueck}}, \bibinfo{journal}{Phys. Rev. Lett.} \textbf{\bibinfo{volume}{38}}, \bibinfo{pages}{347} (\bibinfo{year}{1977}).

\bibitem[{\citenamefont{Miyamoto et~al.}(2017)\citenamefont{Miyamoto, Hara, Hiraki, Masuda, Sasao, Uetake, Yoshimi, Yoshimura, and Yoshimura}}]{miyamoto2017vibrational}
\bibinfo{author}{\bibfnamefont{Y.}~\bibnamefont{Miyamoto}}, \bibinfo{author}{\bibfnamefont{H.}~\bibnamefont{Hara}}, \bibinfo{author}{\bibfnamefont{T.}~\bibnamefont{Hiraki}}, \bibinfo{author}{\bibfnamefont{T.}~\bibnamefont{Masuda}}, \bibinfo{author}{\bibfnamefont{N.}~\bibnamefont{Sasao}}, \bibinfo{author}{\bibfnamefont{S.}~\bibnamefont{Uetake}}, \bibinfo{author}{\bibfnamefont{A.}~\bibnamefont{Yoshimi}}, \bibinfo{author}{\bibfnamefont{K.}~\bibnamefont{Yoshimura}}, \bibnamefont{and} \bibinfo{author}{\bibfnamefont{M.}~\bibnamefont{Yoshimura}}, \bibinfo{journal}{J. Phys. B: At. Mol. Opt. Phys.} \textbf{\bibinfo{volume}{51}}, \bibinfo{pages}{015401} (\bibinfo{year}{2017}).

\bibitem[{\citenamefont{Kinder et~al.}(2021)\citenamefont{Kinder, Cipura, and Halfmann}}]{kinder2021detection}
\bibinfo{author}{\bibfnamefont{J.~F.} \bibnamefont{Kinder}}, \bibinfo{author}{\bibfnamefont{F.}~\bibnamefont{Cipura}}, \bibnamefont{and} \bibinfo{author}{\bibfnamefont{T.}~\bibnamefont{Halfmann}}, \bibinfo{journal}{Phys. Rev. A} \textbf{\bibinfo{volume}{103}}, \bibinfo{pages}{052808} (\bibinfo{year}{2021}).

\bibitem[{\citenamefont{Aharonovich et~al.}(2011)\citenamefont{Aharonovich, Greentree, and Prawer}}]{aharonovich2011diamond}
\bibinfo{author}{\bibfnamefont{I.}~\bibnamefont{Aharonovich}}, \bibinfo{author}{\bibfnamefont{A.~D.} \bibnamefont{Greentree}}, \bibnamefont{and} \bibinfo{author}{\bibfnamefont{S.}~\bibnamefont{Prawer}}, \bibinfo{journal}{Nat. Photonics} \textbf{\bibinfo{volume}{5}}, \bibinfo{pages}{397} (\bibinfo{year}{2011}).

\bibitem[{\citenamefont{Hausmann et~al.}(2014)\citenamefont{Hausmann, Bulu, Venkataraman, Deotare, and Lon{\v{c}}ar}}]{hausmann2014diamond}
\bibinfo{author}{\bibfnamefont{B.}~\bibnamefont{Hausmann}}, \bibinfo{author}{\bibfnamefont{I.}~\bibnamefont{Bulu}}, \bibinfo{author}{\bibfnamefont{V.}~\bibnamefont{Venkataraman}}, \bibinfo{author}{\bibfnamefont{P.}~\bibnamefont{Deotare}}, \bibnamefont{and} \bibinfo{author}{\bibfnamefont{M.}~\bibnamefont{Lon{\v{c}}ar}}, \bibinfo{journal}{Nat. Photonics} \textbf{\bibinfo{volume}{8}}, \bibinfo{pages}{369} (\bibinfo{year}{2014}).

\bibitem[{\citenamefont{Janitz et~al.}(2020)\citenamefont{Janitz, Bhaskar, and Childress}}]{janitz2020cavity}
\bibinfo{author}{\bibfnamefont{E.}~\bibnamefont{Janitz}}, \bibinfo{author}{\bibfnamefont{M.~K.} \bibnamefont{Bhaskar}}, \bibnamefont{and} \bibinfo{author}{\bibfnamefont{L.}~\bibnamefont{Childress}}, \bibinfo{journal}{Optica} \textbf{\bibinfo{volume}{7}}, \bibinfo{pages}{1232} (\bibinfo{year}{2020}).

\bibitem[{\citenamefont{Abulikemu et~al.}(2021)\citenamefont{Abulikemu, Kainuma, An, and Hase}}]{abulikemu2021second}
\bibinfo{author}{\bibfnamefont{A.}~\bibnamefont{Abulikemu}}, \bibinfo{author}{\bibfnamefont{Y.}~\bibnamefont{Kainuma}}, \bibinfo{author}{\bibfnamefont{T.}~\bibnamefont{An}}, \bibnamefont{and} \bibinfo{author}{\bibfnamefont{M.}~\bibnamefont{Hase}}, \bibinfo{journal}{ACS Photonics} \textbf{\bibinfo{volume}{8}}, \bibinfo{pages}{988} (\bibinfo{year}{2021}).

\bibitem[{\citenamefont{Ishioka et~al.}(2006)\citenamefont{Ishioka, Hase, Kitajima, and Petek}}]{ishioka2006coherent}
\bibinfo{author}{\bibfnamefont{K.}~\bibnamefont{Ishioka}}, \bibinfo{author}{\bibfnamefont{M.}~\bibnamefont{Hase}}, \bibinfo{author}{\bibfnamefont{M.}~\bibnamefont{Kitajima}}, \bibnamefont{and} \bibinfo{author}{\bibfnamefont{H.}~\bibnamefont{Petek}}, \bibinfo{journal}{Appl. Phys. Lett.} \textbf{\bibinfo{volume}{89}}, \bibinfo{pages}{231916} (\bibinfo{year}{2006}).

\bibitem[{\citenamefont{Solin and Ramdas}(1970)}]{solin1970raman}
\bibinfo{author}{\bibfnamefont{S.}~\bibnamefont{Solin}} \bibnamefont{and} \bibinfo{author}{\bibfnamefont{A.}~\bibnamefont{Ramdas}}, \bibinfo{journal}{Phys. Rev. B} \textbf{\bibinfo{volume}{1}}, \bibinfo{pages}{1687} (\bibinfo{year}{1970}).

\bibitem[{\citenamefont{Levenson et~al.}(1972)\citenamefont{Levenson, Flytzanis, and Bloembergen}}]{levenson1972interference}
\bibinfo{author}{\bibfnamefont{M.}~\bibnamefont{Levenson}}, \bibinfo{author}{\bibfnamefont{C.}~\bibnamefont{Flytzanis}}, \bibnamefont{and} \bibinfo{author}{\bibfnamefont{N.}~\bibnamefont{Bloembergen}}, \bibinfo{journal}{Phys. Rev. B} \textbf{\bibinfo{volume}{6}}, \bibinfo{pages}{3962} (\bibinfo{year}{1972}).

\bibitem[{\citenamefont{Suchowski et~al.}(2013)\citenamefont{Suchowski, Krogen, Huang, K{\"a}rtner, and Moses}}]{suchowski2013octave}
\bibinfo{author}{\bibfnamefont{H.}~\bibnamefont{Suchowski}}, \bibinfo{author}{\bibfnamefont{P.~R.} \bibnamefont{Krogen}}, \bibinfo{author}{\bibfnamefont{S.-W.} \bibnamefont{Huang}}, \bibinfo{author}{\bibfnamefont{F.~X.} \bibnamefont{K{\"a}rtner}}, \bibnamefont{and} \bibinfo{author}{\bibfnamefont{J.}~\bibnamefont{Moses}}, \bibinfo{journal}{Opt. Express} \textbf{\bibinfo{volume}{21}}, \bibinfo{pages}{28892} (\bibinfo{year}{2013}).

\bibitem[{\citenamefont{Krogen et~al.}(2017)\citenamefont{Krogen, Suchowski, Liang, Flemens, Hong, K{\"a}rtner, and Moses}}]{krogen2017generation}
\bibinfo{author}{\bibfnamefont{P.}~\bibnamefont{Krogen}}, \bibinfo{author}{\bibfnamefont{H.}~\bibnamefont{Suchowski}}, \bibinfo{author}{\bibfnamefont{H.}~\bibnamefont{Liang}}, \bibinfo{author}{\bibfnamefont{N.}~\bibnamefont{Flemens}}, \bibinfo{author}{\bibfnamefont{K.-H.} \bibnamefont{Hong}}, \bibinfo{author}{\bibfnamefont{F.~X.} \bibnamefont{K{\"a}rtner}}, \bibnamefont{and} \bibinfo{author}{\bibfnamefont{J.}~\bibnamefont{Moses}}, \bibinfo{journal}{Nat. Photonics} \textbf{\bibinfo{volume}{11}}, \bibinfo{pages}{222} (\bibinfo{year}{2017}).

\bibitem[{\citenamefont{Chang et~al.}(2021)\citenamefont{Chang, Zheng, Flemens, Heberle, and Moses}}]{chang2021flexible}
\bibinfo{author}{\bibfnamefont{W.-Z.} \bibnamefont{Chang}}, \bibinfo{author}{\bibfnamefont{J.}~\bibnamefont{Zheng}}, \bibinfo{author}{\bibfnamefont{N.}~\bibnamefont{Flemens}}, \bibinfo{author}{\bibfnamefont{D.}~\bibnamefont{Heberle}}, \bibnamefont{and} \bibinfo{author}{\bibfnamefont{J.}~\bibnamefont{Moses}}, in \emph{\bibinfo{booktitle}{Conference on Lasers and Electro-Optics}} (\bibinfo{publisher}{Optica Publishing Group}, \bibinfo{year}{2021}), p. \bibinfo{pages}{SW3Q.5}.

\bibitem[{\citenamefont{Levenson and Bloembergen}(1974)}]{levenson1974dispersion}
\bibinfo{author}{\bibfnamefont{M.}~\bibnamefont{Levenson}} \bibnamefont{and} \bibinfo{author}{\bibfnamefont{N.}~\bibnamefont{Bloembergen}}, \bibinfo{journal}{Phys. Rev. B} \textbf{\bibinfo{volume}{10}}, \bibinfo{pages}{4447} (\bibinfo{year}{1974}).

\end{thebibliography}

\end{document}


\title{Supplemental Materials: Phonon-Mediated Third-Harmonic Generation in Diamond}
\author{Jiaoyang Zheng}
\affiliation{School of Applied and Engineering Physics, Cornell University, Ithaca, NY 14853, USA}
\author{Guru Khalsa}%
\affiliation{Department of Materials Science and Engineering, Cornell University, Ithaca, NY 14853, USA}%
\affiliation{Department of Physics, University of North Texas, Denton, TX 76203, USA}%
\author{Jeffrey Moses}
\affiliation{School of Applied and Engineering Physics, Cornell University, Ithaca, NY 14853, USA}%
\date{\today}
\maketitle

\section{\romannumeral1. Derivation of Phonon-mediated THG and Two-Photon Absorption, Comparison to Coherent Anti-Stokes Raman Scattering and Stimulated Raman scattering}

\begin{figure}
\includegraphics[width=14 cm]
{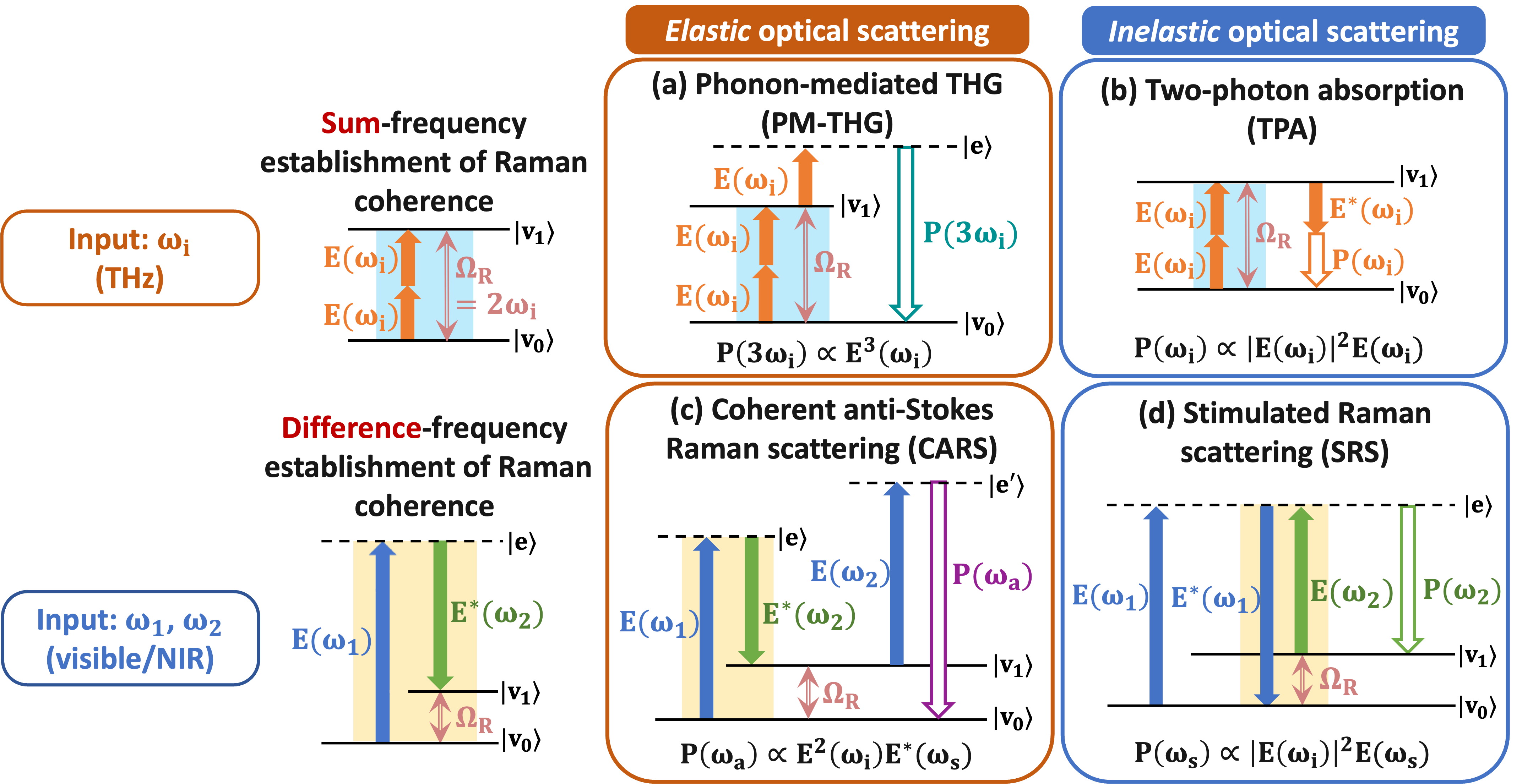}
\caption{\label{fig:table} 
Comparison of Raman coherence establishment and associated optical effects via sum-frequency and difference-frequency pathways. Top row: Sum-frequency driving of Raman coherence using a single THz incident field at $\omega_i = \Omega_R/2$, leading to (a) an elastic optical scattering effect, PM-THG, and (b) an inelastic optical scattering effect, TPA. Bottom row: Conventional difference-frequency driving of Raman coherence using two high-frequency light fields at $\omega_1$ and $\omega_2 =\omega_1-\Omega_R$, resulting in (c) an elastic optical scattering effect, CARS, and (d) an inelastic optical scattering effect, SRS. $\ket{v_0}$, $\ket{v_1}$ are ground and excited states of a Raman phonon. $\ket{e}$, $\ket{e'}$ are virtual electronic excited states.
}
\end{figure}

When the pump field $\vec E$ is polarized along $\theta = 45$° (the [110] axis of crystal), the phonon-mediated third-order polarization, $\vec P_R$, as defined in Eq.~1 of the main text, is parallel to $\vec E$  (refer to Section \uppercase\expandafter{\romannumeral3} for detailed analysis of polarization direction dependence). Therefore, in this section, we write $\vec E$ and $\vec P_R^{(3)}$ in scalar form to simplify the analysis,
\begin{equation}
P_R^{(3)}(t) = \Pi_{0} Q_{R}(t) E(t).
\label{eq:P3R}
\end{equation}

We assume E(t) is monochromatic light in the THz range with frequency $\omega_i = \Omega_R/2$, $E(t) = E(\omega_i) e^{-i\omega_i t} + c.c.$.
The Raman phonon $Q_R$ is driven by two force terms with the sum-beat and difference-beat frequencies of the pump field respectively:
\begin{equation}
    \begin{split}
        M\left(\ddot{Q}_R + 2 \Gamma \dot{Q}_R+\Omega_R^{2}  Q_R\right)
        &= \Pi_{0} E(t) E(t) \\
        &= \Pi_{0} \left(E^2(\omega_i) e^{-i2\omega_i t} + |E(\omega_i)|^2 + c.c. \right)
        \label{eq:eom-S3}
    \end{split}
\end{equation}

The first term, with a frequency of $2\omega_i = \Omega_R$, resonantly drives the Raman phonons. The second term represents a unidirectional force driving at zero frequency, which can be disregarded here due to its contribution being overshadowed by the resonant term.
Therefore, the steady-state solution of Eq.~\ref{eq:eom-S3} is $Q_R(t) = Q_R(2\omega_i) e^{-i2\omega_i t} + c.c.,$
where
\begin{equation}
Q_R(2\omega_i) = \frac{\Pi_{0}}{M\left[\Omega_R^2 - (2\omega_i)^2 - 2i\Gamma(2\omega_i)\right]}E^2(\omega_i).
\label{eq:QR_2omegai}
\end{equation}

By substituting $Q_R$ into Eq.~\ref{eq:P3R}, we obtain two frequency components of $P_R^{(3)}(t)$, one at $3\omega_i$ and the other at the pump frequency, $\omega_i$. 

\begin{equation}
P_R^{(3)}(t) = P_R^{(3)}(3\omega_i = \omega_i+\omega_i+\omega_i) e^{-i3\omega_i t} + P_R^{(3)}(\omega_i = \omega_i+\omega_i-\omega_i) e^{-i\omega_i t} + c.c.
\label{eq:PR3-S3}
\end{equation}
where
\begin{equation}
P_R^{(3)}(3\omega_i = \omega_i+\omega_i+\omega_i)  = \Pi_0 Q_R(2\omega_i)E(\omega_i) = \chi_R^{(3)}(2\omega_i\simeq \Omega_R)E(\omega_i)E(\omega_i)E(\omega_i)
\label{eq:PR3omega}
\end{equation}
represents the origin of phonon-mediated third-harmonic generation (PM-THG), with the Raman susceptibility given by
\begin{equation}
\chi_R^{(3)}(2\omega_i \simeq \Omega_R) = \frac{\Pi_0^2}{M \left[\Omega_R^2-(2\omega_i)^2-2i \Gamma (2\omega_i) \right]}.
\label{eq:chi_R_THG}
\end{equation}
The other polarization term at the incident frequency is
\begin{equation}
P_R^{(3)}(\omega_i = \omega_i+\omega_i-\omega_i) = \Pi_0 Q_R(2\omega_i)E^*(\omega_i) = \chi_R^{(3)}(2\omega_i\simeq \Omega_R)\left|E(\omega_i)\right|^2 E(\omega_i),
\label{eq:PR1omega}
\end{equation}
which describes two-photon absorption (TPA) of the pump field $E(t)$. Raman susceptibility $\chi_R^{(3)}(2\omega_i\simeq \Omega_R)$ is positive imaginary at the Raman resonance (see Fig.~\ref{fig:susceptibility}), resulting in nonlinear absorption of $E(t)$ at a rate proportional to its intensity $\left|E(\omega_i)\right|^2$. Further details on TPA are provided in Ref.~\cite{agrawal2000nonlinear}.



\begin{table}
\centering
\renewcommand{\arraystretch}{1.5}
\begin{tabular}
{|P{1.5cm}|P{1.3cm}|P{2.5cm}|P{2.8cm}|P{1.5cm}|P{7cm}|  }
\hline
Process&Input beam&Frequency range&Sum/Difference-frequency pathway& Elastic / Inelastic &$P^{(3)}$
\\
\hline
PM-THG & $\omega_i$ & THz ($\approx\Omega_R/2$) & sum & elastic & $P_R^{(3)}(3\omega_i)  = \chi_R^{(3)}(2\omega_i\simeq\Omega_R)E^3(\omega_i)$ \\
\hline
TPA & $\omega_i$ &THz ($\approx \Omega_R/2$) & sum & inelastic & $P_R^{(3)}(\omega_i) = \chi_R^{(3)}(2\omega_i\simeq\Omega_R)\left|E(\omega_i)\right|^2 E(\omega_i)$ \\
\hline
CARS & $\omega_p$, $\omega_s$ & visible/near-IR ($\gg\Omega_R$) & difference & elastic & $P_R^{(3)}(\omega_a) = \chi_R^{(3)}(\omega_p-\omega_s\simeq\Omega_R)E^2(\omega_p)E^*(\omega_s)$ \\
\hline
SRS & $\omega_p$, $\omega_s$  & visible/near-IR ($\gg\Omega_R$) & difference & inelastic & $P_R^{(3)}(\omega_s)  = \chi_R^{(3)}(-\omega_p+\omega_s\simeq -\Omega_R)\left|E(\omega_p)\right|^2 E(\omega_s)$\\
\hline
\end{tabular}
\caption{\label{table} Summary of characteristics of PM-THG, TPA, CARS, and SRS.}
\end{table}

The above model can also be applied to analyze stimulated Raman scattering (SRS) and coherent anti-Stokes Raman scattering (CARS), both of which typically require two visible or near-IR laser beams, a pump beam $E(\omega_p)e^{-i\omega_i t} +c.c.$, and a Stokes beam $E(\omega_s)e^{-i\omega_s t}+c.c.$ with $\omega_s = \omega_p-\Omega_R$. The Raman phonon is resonantly driven by the difference-beat frequency, such that $Q_R(t) = Q_R(\omega_p-\omega_s) e^{-i(\omega_p-\omega_s) t} + c.c.$,
where
\begin{equation}
Q_R(\omega_p-\omega_s) = \frac{\Pi_{0}}{M\left[\Omega_R^2 - (\omega_p-\omega_s)^2 - 2i\Gamma(\omega_p-\omega_s)\right]}E(\omega_p)E^*(\omega_s).
\label{eq:QR_omega_p_omega_s}
\end{equation}

Substituting $Q_R(t)$ into Eq.~\ref{eq:P3R} , we derive one third-order polarization term at the anti-Stokes frequency $\omega_a=\omega_p+\Omega_R$, and another at the Stokes frequency $\omega_s$:
\begin{equation}
P_R^{(3)}(\omega_a = \omega_p-\omega_s+\omega_p) = \Pi_0 Q_R(\omega_p-\omega_s)E(\omega_p) = \chi_R^{(3)}(\omega_p-\omega_s\simeq \Omega_R)E^2(\omega_p)E^*(\omega_s)
\label{eq:PR_CARS}
\end{equation}
describes the origin of CARS, with 
\begin{equation}
\chi_R^{(3)}(\omega_p-\omega_s\simeq \Omega_R) = \frac{\Pi_0^2}{M \left[\Omega_R^2-(\omega_p-\omega_s)^2-2i \Gamma (\omega_p-\omega_s) \right]}. 
\label{eq:chi_R_CARS}
\end{equation}
Compared to Eq.~\ref{eq:PR3omega}, both CARS and PM-THG involve driving Raman phonon coherence as an intermediate step, albeit through different routes — the former via a difference-frequency pathway, and the latter via a sum-frequency pathway. As Fig.~\ref{fig:table} shows, CARS and PM-THG are both elastic four-wave mixing (FWM) processes, where the Raman coherence increases the optical scattering cross-section but all incident energy is returned to the optical fields. Therefore, we regard PM-THG as a low-frequency, single-laser-excitation variant of CARS. 


The nonlinear polarization term at the Stokes frequency is
\begin{equation}
P_R^{(3)}(\omega_s = -\omega_p+\omega_s+\omega_p)  = \Pi_0 Q_R(-\omega_p+\omega_s)E(\omega_p) = \chi_R^{(3)}(-\omega_p+\omega_s\simeq -\Omega_R)\left|E(\omega_p)\right|^2 E(\omega_s),
\label{eq:PR_SRS}
\end{equation}
which describes the SRS process. The corresponding Raman susceptibility, $\chi_R^{(3)}(-\omega_p+\omega_s\simeq -\Omega_R)$, is negative imaginary, suggesting an exponential growth of the Stokes field at a rate proportional to the pump intensity $\left|E(\omega_p)\right|^2$ \cite{boyd2020nonlinear}.  Similar to TPA (Eq.~\ref{eq:PR1omega}), the optical field energy is absorbed to drive lattice vibrations, and thus not conserved in SRS. Therefore, as illustrated in Fig.~\ref{fig:table} (a) and (c), SRS and TPA are inelastic optical processes that accompany Raman coherence establishment via difference-frequency and sum-frequency pathway, respectivelys. Phase matching is automatically fulfilled in these two processes.

The relationships and features of the four processes (PM-THG, TPA, CARS, and SRS) are summarized in Fig.~\ref{fig:table} and Table S2.


\section{\romannumeral2. Raman Microscopy Measurements}
\begin{figure*}[ht]
  \includegraphics[width=8 cm]{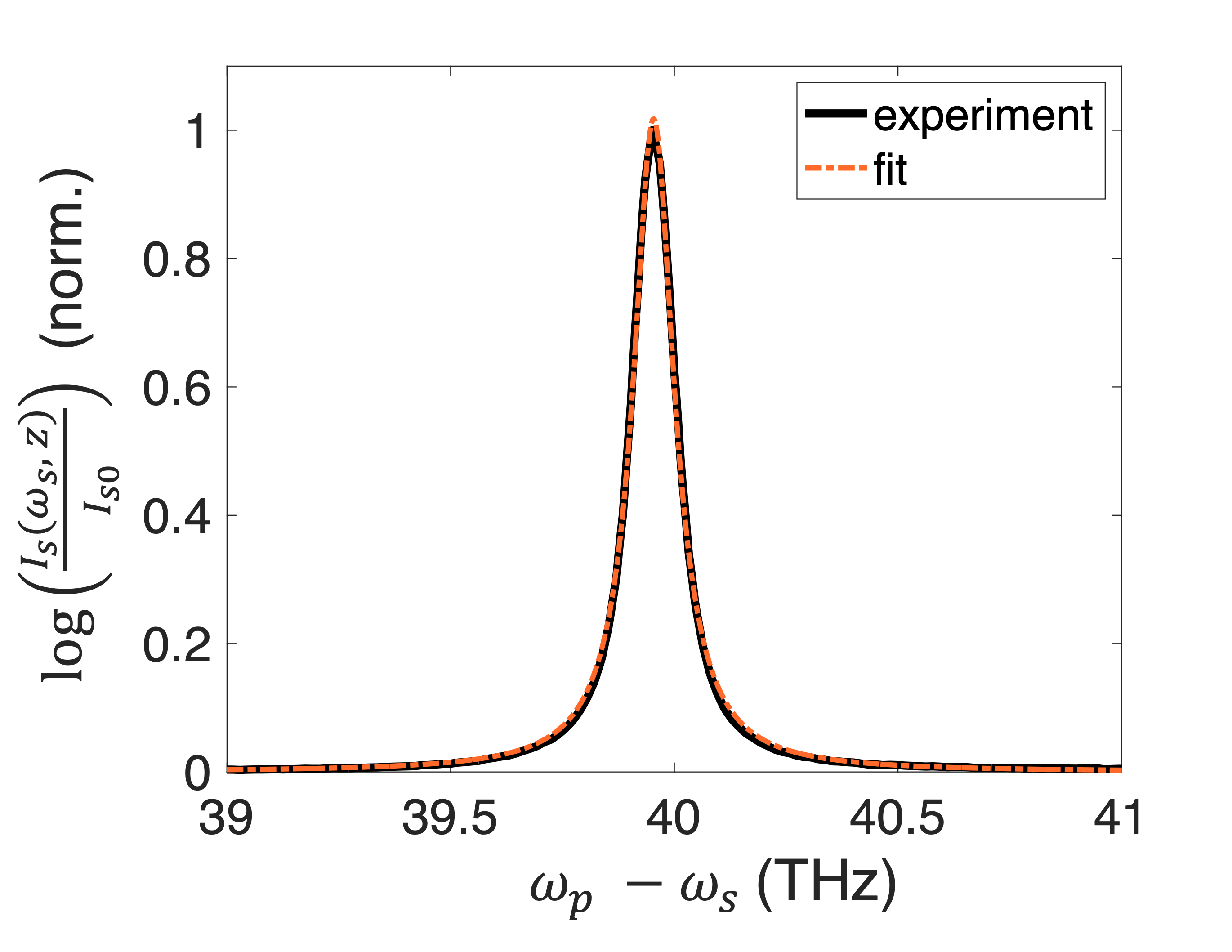}
  \caption{The measured intensity spectrum of the Raman signal after a log transformation (black) and the fitted curve (orange).
  \label{Raman}}
\end{figure*}

Figure \ref{Raman} shows the spectral intensity of the spontaneous Raman signal generated from the diamond sample. This signal was acquired through transmission measurements using a WITec Alpha300R confocal Raman microscope equipped with a continuous wave (CW) laser operating at a 532 nm wavelength.

From Ref.~\cite{boyd2020nonlinear}, the evolution of the Raman signal amplitude $A_s(\omega_s,z)$ propagating along the z-axis (the [001] axis of crystal) is
\begin{equation}\label{dAsdz}
\frac{d A_s \left(\omega_s,z \right)}{d z} = -\alpha \left(\omega_s \right)A_s \left(\omega_s,z \right)
\end{equation}
within the slowly varying amplitude approximation. Here, $z$ is the propagation distance, and $\omega_s$ represents the Raman signal frequency. $\alpha \left(\omega_s \right)$ is defined as an absorption coefficient \cite{boyd2020nonlinear},
\begin{equation}\label{alpha}
\alpha \left(\omega_s \right) = -i\frac{\omega_s}{2\epsilon_0 n \left(\omega_s \right) c} \chi_R^{(3)} \left(-\omega_p+\omega_s \right) |A_p|^2,
\end{equation}
where $\epsilon_0$ is the vacuum permittivity, $n\left(\omega_s \right)$ is the linear refractive index, and c is the speed of light. $A_p$ represents the electric field amplitude of the pump at frequency $\omega_p$, which is treated as a constant with no depletion here due to the low efficiency of the spontaneous Raman scattering. $\chi_R^{(3)}$ is the Raman susceptibility given by Eq. \ref{eq:chi_R_CARS}.

The solution of Eq.~\ref{dAsdz} is
\begin{equation}\label{As}
A_s \left(\omega_s,z \right)= A_s \left(\omega_s,0 \right)e^{-\alpha \left(\omega_s \right)z},
\end{equation}
and thus the intensity of the Raman signal is 
\begin{equation}\label{Is}
I_s \left(\omega_s,z \right) =  2\epsilon_0 n(\omega_s) c \left| A_s\left(\omega_s,z\right) \right|^2 = I_{s0}e^{-2 \operatorname{Re}(\alpha\left(\omega_s\right))z}.
\end{equation}

Since there is only one incident beam at $\omega_p$, the input Raman signal $I_s \left(\omega_s,z\right)$ at $z=0$ arises solely from quantum noise, and thus can be treated as a constant $I_{s0}$ 
over a narrow frequency band near the Stokes frequency, $\omega_p-\Omega_R$.


After a log transformation, Eq.~\ref{Is} becomes 
\begin{equation}\label{log_Is}
\log\left(\frac{I_s \left(\omega_s,z \right)}{I_{s0}}\right) =   -2 \operatorname{Re}(\alpha\left(\omega_s\right))z = \beta \operatorname{Re}\left(\frac{i \omega_s }{\Omega_R^2 - \left(\omega_p-\omega_s \right)^2 + 2i\Gamma \left(\omega_p-\omega_s \right)}\right),
\end{equation}
where $\Omega_R$ represents the Raman resonant frequency, and $\Gamma$ is the phonon line-width (half-width at half-maximum (HWHM)).


We take the baseline of the measured spectral intensities of the Raman signal (the intensity at frequencies several $\Gamma$ away from the Stokes frequency) as the value of $I_{s0}$. 
$\beta \left(\equiv \frac{\Pi_0^2 z}{\epsilon_0 n(\omega_s)cM} |A_p|^2 \right)$ 
is a constant to be fitted, where the dispersion of the refractive index $n(\omega_s)$ is neglected.

The black curve in Fig.~\ref{Raman} represents the measured spectral intensities of the Raman signal, $I_s \left(\omega_s,z \right)$, after applying the log transformation described above, and is plotted against the frequency shift ($\omega_p-\omega_s$). A least-squares fitting of the parameters $\beta$, $\Omega_R$, and $\Gamma$ in Eq. \ref{log_Is} yields the orange curve, which closely matches the experimental data.

Table S1 compares our fitted values of $f_R (=\Omega_R/(2\pi))$ and $\gamma (=\Gamma/(2\pi))$ with those reported in the literature for the F$_{2g}$ Raman phonon in diamond \cite{maehrlein2017terahertz,solin1970raman,ishioka2006coherent,levenson1972interference,laubereau1971decay}. Our $f_R$ is consistent with previously reported values, while our $\gamma$ falls on the higher side of a range of reported values that vary by a factor of $\sim$2.4. This variation is expected as the line width depends on the crystal quality and preparation.

\begin{table}
\centering
\renewcommand{\arraystretch}{1.5}
\begin{tabular}{|P{4.3cm}|P{2cm}|P{2.1cm}|P{1cm}|P{2.1cm}|P{1cm}|P{2.1cm}|P{2.1cm}|  }
\hline
References & \cite{maehrlein2017terahertz}$^\circ$ & \cite{solin1970raman}$^\star$ & \cite{ishioka2006coherent}$^{\star}$ & \cite{ishioka2006coherent}$^{\circ}$ & \cite{levenson1972interference}$^\star$ & \cite{laubereau1971decay}$^\circ$ & Our results
\\
\hline
Resonant frequency $f_R$ (THz) & $39.95\pm0.01$ & $39.95\pm0.015$ & 39.91 & $39.84\pm0.03$ & 39.93 & $\backslash$ & $39.952\pm 0.002$\\
\hline
Line width $\gamma$ (THz)  & 0.045 & $0.0495\pm0.0006$ & 0.023 & $0.023\pm0.006$ & 0.031 &$0.055\pm0.006$ & $0.056 \pm 0.003$\\
\hline
\end{tabular}
\caption{\label{table} The resonant frequency ($f_R$) and the HWHM line-width ($\gamma$) for the $F_{\text{2g}}$ phonon of diamond, as reported in prior literature and as observed in our experiments. $^{\star}$ represents the parameters obtained in the frequency domain. $^{\circ}$ represents the parameters obtained in the time domain, wherein $\gamma$ is calculated as the phonon decay rate (in units of \si{ps^{-1}}) from \cite{maehrlein2017terahertz,ishioka2006coherent,laubereau1971decay}, divided by $2\pi$.}
\end{table}

\section{\romannumeral3. Pump Polarization dependence analysis of THG signal intensity}
This section analyzes the pump polarization dependence of the PM-THG signal and the pure electronic THG signal by considering the vectorial nature of the pump field $\vec{E}$ and third-order polarization field $\vec{P}^{(3)}$.

Assuming the THz pump field, $\vec E(t) = \vec E(\omega_0)e^{-i\omega_0 t} + c.c.$ with $\omega_0 = \Omega_R/2$, is polarized along the $\theta$ direction within the x-y plane of the crystal, such that
\begin{equation}
\vec E(\omega_0) = E_0 \left[\begin{matrix}\cos \theta  & \sin \theta & 0\end{matrix}\right],
\label{eq:E}
\end{equation}
where x, y, and z correspond to the [100], [010], and [001] crystalline axes respectively, and $E_0$ represents the electric field amplitude.

From Eq.~1 in the main text, the phonon-mediated contribution to the third-order polarization $\vec P^{(3)}$ at frequency of $3\omega_0 $ is
\begin{equation}
P_{R,i}^{(3)}(3\omega_0)  = \sum_{j}\Pi_{ij}Q_R E_j(\omega_0)
\label{eq:PR}.
\end{equation}

From Eq.~2 and the tensor form of Raman polarizability $\Pi$ provided in the main text, the coherent phonon amplitude is
\begin{equation}
    \begin{split}
    Q_R &= \frac{1}{M\left[\Omega_R^2 - (2\omega_0)^2 - 2i\Gamma(2\omega_0)\right]}\sum_{ij} \Pi_{ij} E_i(\omega_0) E_j(\omega_0) \\
    & = \frac{\Pi_{0}}{M\left[\Omega_R^2 - (2\omega_0)^2 - 2i\Gamma(2\omega_0)\right]}E_0^2\sin\left(2\theta\right),
    \end{split}
\label{eq:Q_R_vec}
\end{equation}
which has $\sin(2\theta)$ dependence for the pump polarization angle, consistent with the analysis in Ref.~\cite{ishioka2006coherent}.

Inserting Eq.~\ref{eq:E} and \ref{eq:Q_R_vec} into Eq.~\ref{eq:PR}, we obtain 
\begin{equation}
\overrightarrow {P_R^{(3)}}(3\omega_0) = \chi_R^{(3)}(2\omega_0)E_0^3\sin\left(2\theta\right) \left[\begin{matrix}\sin \theta  & \cos \theta & 0\end{matrix}\right].
\label{eq:PR_vec}
\end{equation}

\begin{figure*}[ht]
  \includegraphics[width=8.5 cm]{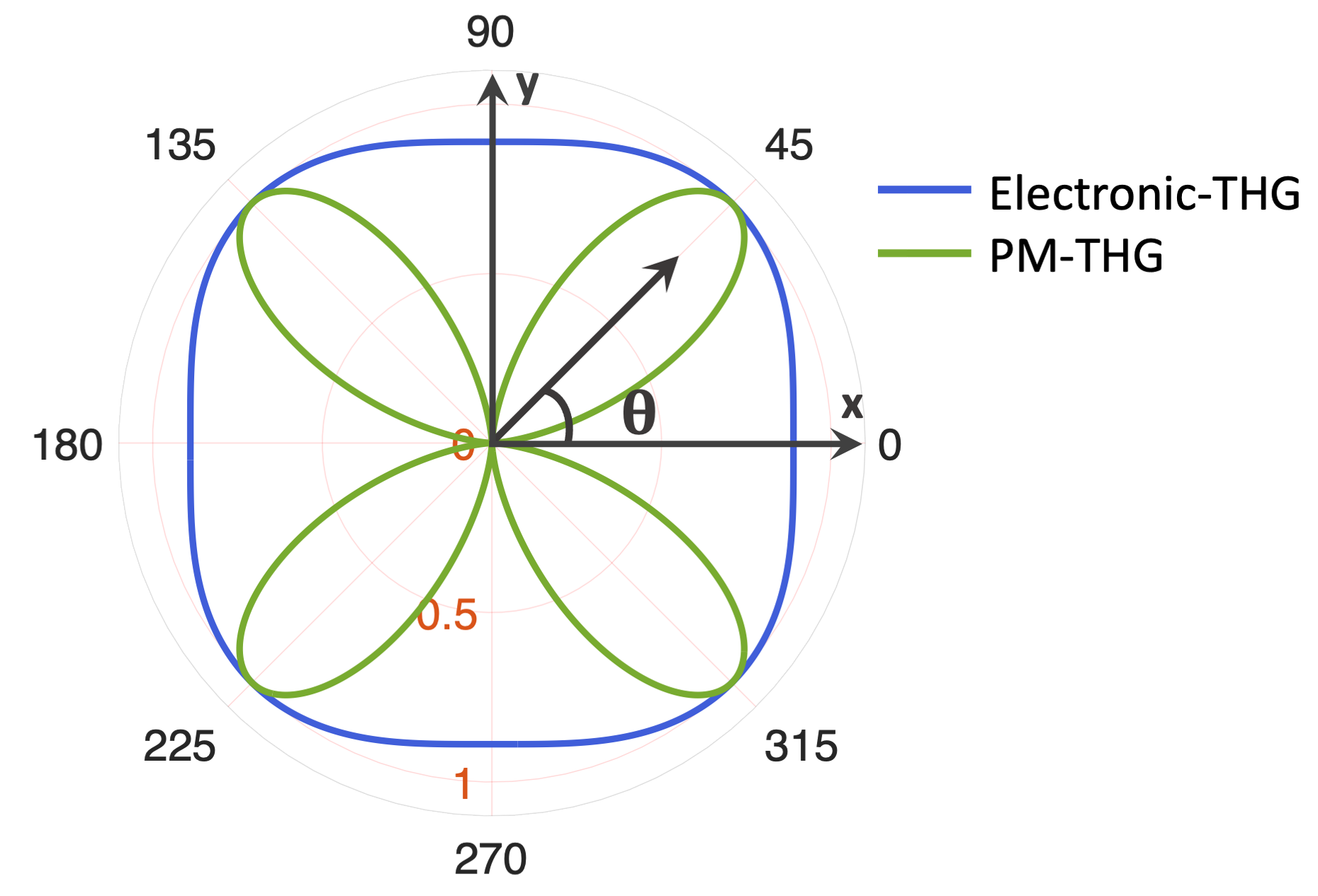}
  \caption{PM-THG (green) and the non-resonant electronic-THG (blue) intensities as a function of the pump polarization angle $\theta$, both normalized to their intensity at $\theta = 45$°.\label{I_3}}
\end{figure*}

Therefore, the intensity of PM-THG
\begin{equation}
I_R(3\omega_0) \propto
\left|\overrightarrow {P_R^{(3)}}(3\omega_0)\right|^2 = \left|\chi_R^{(3)}(2\omega_0)\right|^2 \left|E_0\right|^6\sin^2\left(2\theta\right),
\label{eq:IR}
\end{equation}
which manifests as a $\sin^2(2\theta)$ dependency on the pump polarization angle, as the green curve in Fig.~\ref{I_3} shows. Such angle dependence is consistent with the conventional Raman signal generated from the same F$_{2g}$ phonon in a diamond crystal using a visible pump \cite{brown1995site}, affirming the link between optical scattering physics in the THz domain through sum-frequency driving of Raman coherence and those through traditional difference-frequency coherence excitation with high-frequency light.

However, the pure electronic THG has a significantly different $\theta$ dependence. From Eq.~1 in the main text, the non-resonant electronic contribution to the third-order polarization $\vec P^{(3)}$ at frequency of $3\omega_0 $ is
\begin{equation}
P_{e,i}^{(3)}(3\omega_0) = \sum_{jkl}\chi_{e,ijkl}^{(3)} E_j(\omega_0) E_k(\omega_0) E_l(\omega_0),
\label{eq:P3e}
\end{equation}

The diamond crystal, belonging to space group 227 (Fd$\bar{3}$m), has two independent components of $\chi_{e,ijkl}^{(3)}$ dictated by Kleinman's symmetry relation \cite{aroyo2011crystallography}: $\chi_{e,1111}^{(3)}$, and $\chi_{e,1221}^{(3)}=\chi_{e,1212}^{(3)}=\chi_{e,1122}^{(3)}$, where the indices 1 and 2 denote two different Cartesian axes. Therefore,
\begin{equation}
\begin{split}
P_{e,x}^{(3)}(3\omega_0) 
&= \chi_{e,xxxx}^{(3)} E_x^3(\omega_0)+ \left(\chi_{e,xyyx}^{(3)}+\chi_{e,xyxy}^{(3)}+\chi_{e,xxyy}^{(3)}\right)E_y^2(\omega_0)E_x(\omega_0)\\
&=\left[\chi_{e,1111}^{(3)} \cos^3(\theta)+ 3\chi_{e,1221}^{(3)}\sin^2(\theta)\cos(\theta)\right]E_0^3,
\end{split}
\label{eq:P3ex}
\end{equation}
and
\begin{equation}
\begin{split}
P_{e,y}^{(3)}(3\omega_0) 
&= \chi_{e,yyyy}^{(3)} E_y^3(\omega_0)+ \left(\chi_{e,yxxy}^{(3)}+\chi_{e,yxyx}^{(3)}+\chi_{e,yyxx}^{(3)}\right)E_x^2(\omega_0)E_y(\omega_0)\\
&=\left[\chi_{e,1111}^{(3)} \sin^3(\theta)+ 3\chi_{e,1221}^{(3)}\cos^2(\theta)\sin(\theta)\right]E_0^3,
\end{split}
\label{eq:P3ey}
\end{equation}

By substituting the susceptibility values from Ref.~\cite{levenson1974dispersion}, $\chi_{e,1111}^{(3)} = 4.60$ ($10^{-14}$ esu) and $\chi_{e,1221}^{(3)} = 1.72$ ($10^{-14}$ esu), into Eq.~\ref{eq:P3ex} and \ref{eq:P3ey}, we calculate the signal intensity of the non-resonant electronic THG, $I_e(3\omega_0) \propto \left|\overrightarrow {P_e^{(3)}}(3\omega_0)\right|^2$, as a function of the polarization angle $\theta$. As the blue curve in Fig.~\ref{I_3} shows, the non-resonant electronic THG is notably less sensitive to the pump polarization angle, with an only 1.12-fold difference between its maximum and minimum intensity. 

\section{\romannumeral4. Derivation of PM-THG and electronic THG for pulsed excitation and calculation of susceptibility ratio}

For simplified analysis, 
we have assumed the THz pump field $E(t)$ to be a monochromatic light in Sec.~\uppercase\expandafter{\romannumeral1} and \uppercase\expandafter{\romannumeral3}. 
While this approach sufficiently illustrates the resonant behavior of PM-THG and its dependency on pump polarization angle, 
it cannot be applied to simulate the THG spectral intensity under excitation of an picosecond laser pump possessing a broad bandwidth.
Furthermore, the susceptibility ratio $\left| \frac{\chi_R^{(3)}(\Omega_R)}{\chi_e^{(3)}} \right|$ cannot be simply determined by the ratio between the square root of the THG signal intensity measured at $\theta = 45$° and that at $\theta = 0$°. Its calculation requires a more careful analysis involving the autoconvolution of the Fourier transform of the pulsed laser field $E(t)$,
\begin{equation}
E(t)=\int \tilde E(\omega) \exp (-i \omega t) d \omega,
\label{E_FT}
\end{equation}
where $\tilde E(\omega)$ is the Fourier transform of $E(t)$, and $|\tilde E(\omega)|^2$ represents the intensity spectrum of the pump pulse. Again, we consider the pump field $\vec E$ polarized along $\theta = 45$° in this section, and write $\vec E$ and the third-order polarization $\vec P^{(3)}$ in scalar form to simplify the analysis. 


From Eqs. 1 and 2 in the main text, the Fourier transform of $P^{(3)}(t)$ is 
\begin{equation}
\tilde P^{(3)}(\omega) = \chi_e^{(3)} \left( \tilde E(\omega) * \tilde E(\omega)\right) *\tilde E(\omega) + \Pi_0 \tilde Q_{R}(\omega) * \tilde E(\omega), 
\label{eq:P3_FT}
\end{equation}
and the transform of $Q_R(t)$ is 
\begin{equation}
M\left( \left(-i\omega\right)^2\tilde Q_R - 2 i\Gamma \omega \tilde Q_R+\Omega_R^{2} \tilde Q_R\right)
= \Pi_0 \tilde E * \tilde E,
\label{eq:eom_FT}
\end{equation}
where $*$ represents linear convolution. The non-resonant third-order electronic susceptibility $\chi_e^{(3)}$ is regarded as a constant with no dispersion in the THz region, because the photon energy is far below the bandgap of diamond (5.5 eV) \cite{levenson1972interference,kozak2012two}.

The analytical solution of Eq.~\ref{eq:eom_FT} is
\begin{equation}
\tilde Q_R(\omega) = \frac{\Pi_0}{M (\Omega_R^2-\omega^2-2i \Gamma \omega )} \left(\tilde E * \tilde E \right).
\label{eq:Q_R}
\end{equation}


By substituting $\tilde Q_R(\omega)$ into Eq.~\ref{eq:P3_FT}, we obtain

\begin{flalign}
\nonumber
\tilde P^{(3)}(\omega) &= \chi_e^{(3)} \left( \tilde E(\omega) * \tilde E(\omega)\right) * \tilde E(\omega) + \left[\frac{\Pi_0^2}{M \left(\Omega_R^2-\omega^2-2i \Gamma \omega \right)}\left(\tilde E(\omega) * \tilde E(\omega) \right) \right]* \tilde E(\omega) \\
\nonumber
&\equiv \left[\left(\chi_e^{(3)} + \chi_R^{(3)}(\omega) \right) \left(\tilde E(\omega) * \tilde E(\omega)\right)\right]* \tilde E(\omega) \\
&\equiv \tilde P_e^{(3)}(\omega) + \tilde P_R^{(3)}(\omega),
\label{eq:P3_FT_total}
\end{flalign}
where $ \tilde P_e^{(3)}(\omega) = \left[\chi_e^{(3)}\left(\tilde E(\omega) * \tilde E(\omega)\right)\right]* \tilde E(\omega)  $ represents the pure electronic contribution to the third-order polarization, and  $ \tilde P_R^{(3)}(\omega) = \left[\chi_R^{(3)}(\omega)\left(\tilde E(\omega) * \tilde E(\omega)\right)\right]* \tilde E(\omega)  $ represents the phonon-mediated pathway contribution. The Raman susceptibility $\chi_R^{(3)}(\omega)$ in this equation is given by
\begin{equation}
\chi_R^{(3)}(\omega) = \frac{\Pi_0^2}{M \left(\Omega_R^2-\omega^2-2i \Gamma \omega \right)} = \chi_e^{(3)} \eta \left(\frac{2\Gamma \Omega_R}{\Omega_R^2-\omega^2-2i \Gamma \omega}\right),
\label{eq:chi_R}
\end{equation}
with an introduced constant $\eta$, defined as
\begin{equation}
\eta =\frac{\Pi_0^2}{M \left(2\Gamma \Omega_R\right)\chi_e^{(3)}}.
\label{eq:R}
\end{equation}

By integrating the measured pump spectra shown in Fig.~2(b) from the main text, alongside the phonon parameters determined in Sec.~\uppercase\expandafter{\romannumeral2},  into Eq.~\ref{eq:P3_FT_total}, we derive the respective signal spectra for each of the PM-THG mechanism ($\left|\tilde P_R^{(3)}(\omega)\right|^2$) and the pure electronic-THG mechanism ($\left|\tilde P_e^{(3)}(\omega)\right|^2$), as shown in Fig.~2(c) and (d) respectively in the main text.

We introduced the constant $\eta$ into Eq.~\ref{eq:chi_R} since it directly defines the ratio between $|\chi_R^{(3)}|$ at the phonon resonance and $|\chi_e^{(3)}|$.  
By equating the THG enhancement factor of 113 measured at resonance to $|(P_R^{(3)}+P_e^{(3)})/P_e^{(3)}|^2$, we can deduce the value of the constant $\eta$ and thus the susceptibility ratio:
\begin{equation}
\eta = \left| \frac{\chi_R^{(3)}(\omega = \Omega_R)}{\chi_e^{(3)}} \right| \geq 58. 
\label{eq:ratio}
\end{equation}
Due to experimental uncertainty in polarization purity, there may be some phonon-mediated contribution to the measured THG signal at $\theta = 0$°, potentially leading to an underestimation of the actual enhancement factor. Therefore, we incorporate this uncertainty by utilizing the inequality notation in this expression.

The susceptibility ratio we obtained is approximately three times the value of $21 (\pm 3)$ reported in Ref.~\cite{levenson1972interference}. This discrepancy can be attributed to different definitions of nonlinear susceptibility for different nonlinear optical processes. Our study investigates a THG process with a purely electronic susceptibility given by $\chi_{e}^{(3)}(3\omega_0; \omega_0, \omega_0, \omega_0) = \chi_{e,1111}^{(3)}+3\chi_{e,1221}^{(3)}$ for a pump field along the [110] direction (see Sec.~\uppercase\expandafter{\romannumeral3}). However, Ref.~\cite{levenson1972interference} examines a FWM process with a purely electronic susceptibility $\chi_{e}^{(3)}(2\omega_1+
\omega_2; \omega_1, \omega_1, -\omega_2) = 3(\chi_{e,1111}^{(3)}+3\chi_{e,1221}^{(3)})$ when two incident light fields at $\omega_1$ and $\omega_2$ are both along the [110] direction. The electronic susceptibility in the FWM process is three times that of the THG process, resulting in a lower susceptibility ratio between $\chi_{R}^{(3)}$ and $\chi_{e}^{(3)}$, as reported in Ref.~\cite{levenson1972interference}. Additionally, our experiment employs a laser in the THz range, while the pump source in Ref.~\cite{levenson1972interference} operates in the visible range with photon energy closer to the bandgap. Variations in the third-order electronic susceptibility across different frequency ranges \cite{almeida2017nonlinear} may also contribute to the observed difference in susceptibility ratios.

\section{\romannumeral5. THG suppression}
\begin{figure*}[ht]
  \includegraphics[width=6.9 cm]{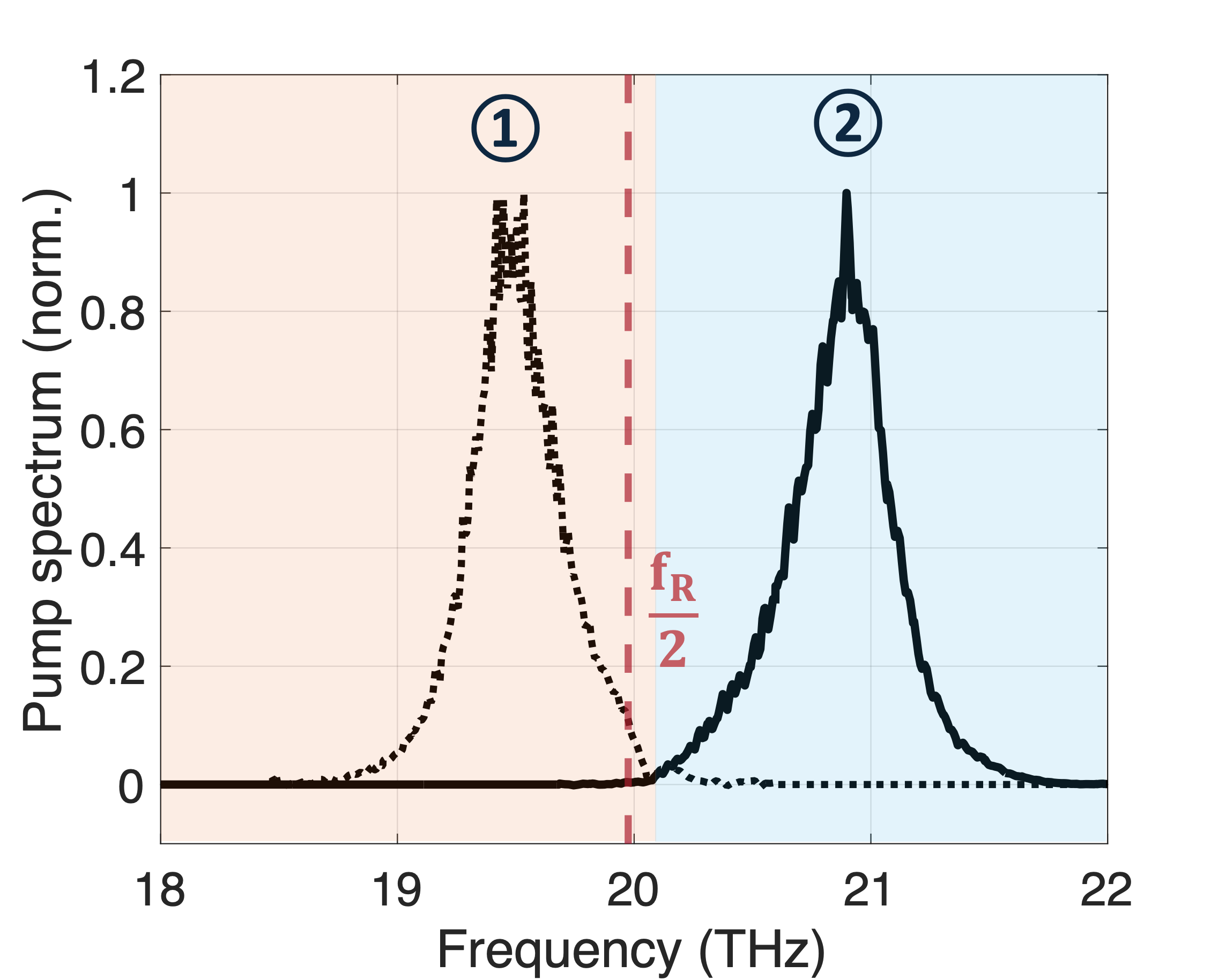}
  \caption{Pump spectra corresponding to the signal displayed in Fig.~3(d) in the main text, \ding{172} blow and \ding{173} above half phonon resonance ($f_R/2$, red dashed line) respectively. \label{detune_pump}}
\end{figure*}
\begin{figure*}[ht]
  \includegraphics[width=9 cm]{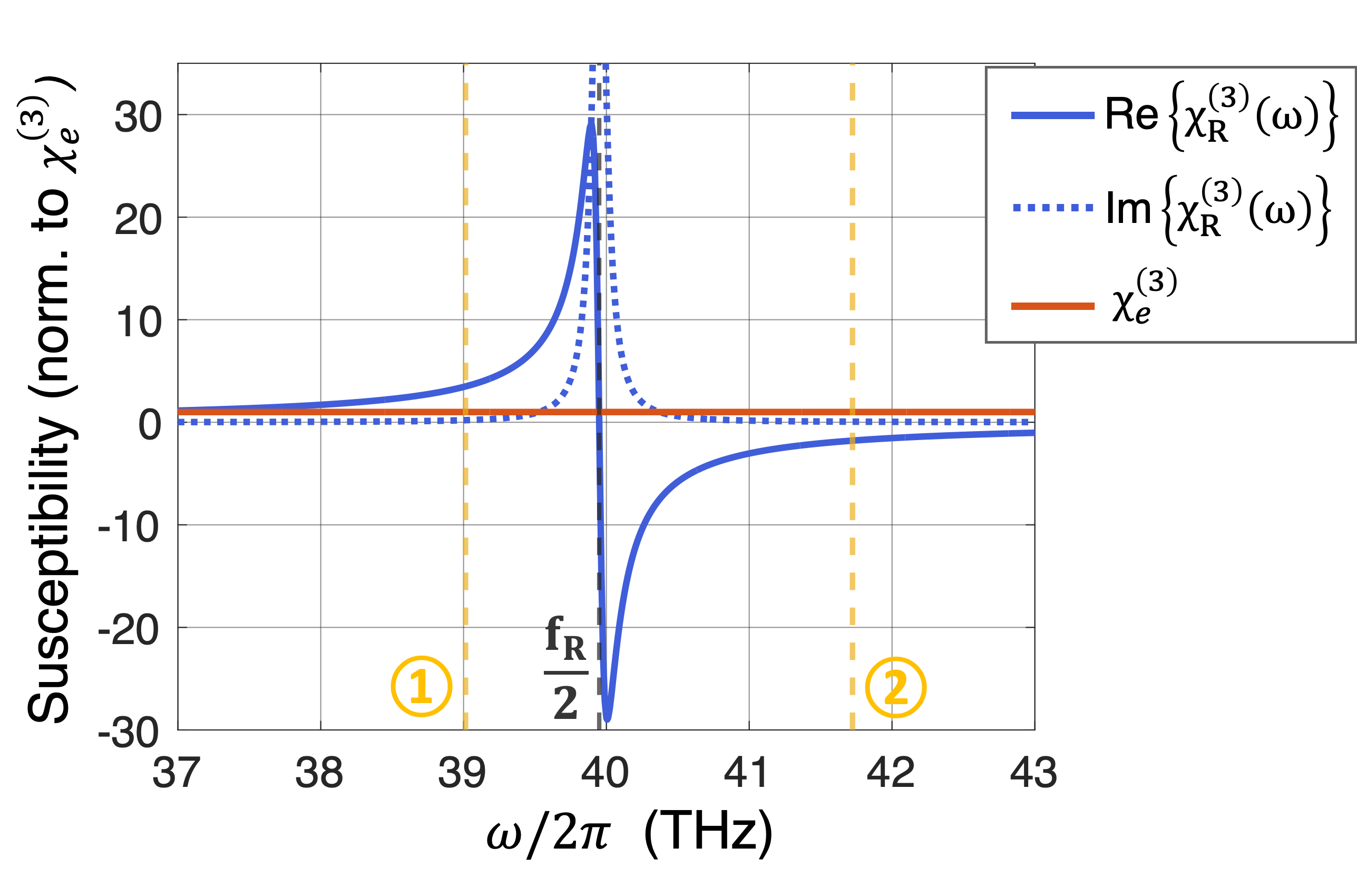}
  \vspace{-10 pt}
  \caption{The real and imaginary part of Raman susceptibility $\chi^{(3)}_R$ (blue) as a function of $\omega$ (Eq.~\ref{eq:chi_R}), normalized to the frequency-independent third-order electronic susceptibility $\chi^{(3)}_e$ (red). The yellow dashed lines represent twice the center frequency of pump \ding{172} and \ding{173} in Fig.~\ref{detune_pump}.
  \label{fig:susceptibility}}
\end{figure*}

Fig.~\ref{detune_pump} presents the pump spectra corresponding to the signal spectra shown in Fig.~3(d) of the main text, \ding{172} below and \ding{173} above resonance respectively.
Fig.~\ref{fig:susceptibility} displays the real and imaginary parts of Raman susceptibility $\chi^{(3)}_R(\omega/2\pi)$, normalized to the frequency-independent electronic susceptibility $\chi^{(3)}_e$.
When the pump has a center frequency $f_i = f_R/2$, the corresponding Raman susceptibility $\chi^{(3)}_R(\omega/2\pi=2f_i)$ is pure imaginary at Raman resonance. As the pump frequency deviates from $f_R/2$, the imaginary part of $\chi^{(3)}_R$ rapidly diminishes to zero, while the real part starts to dominate the generation of the PM-THG signal. 
For instance, the center frequency of pump \ding{172} in Fig.~\ref{detune_pump} corresponds to a Raman susceptibility $\chi^{(3)}_R(\omega)$ whose real part is much higher than its imaginary part, marked by the yellow dashed line \ding{172} in Fig.~\ref{fig:susceptibility}. 
Since the real part of $\chi^{(3)}_R(\omega)$ has the same sign as $\chi^{(3)}_e$, the phonon-mediated and electronic contributions to the polarization field combine constructively and thus enhance the total THG efficiency, as observed in Fig.~3(d) of the main text.  However, the enhancement factor at this pump frequency is significantly smaller than at resonance.

Conversely, when the pump frequency $f_i$ is tuned above $f_R/2$, the real part of $\chi^{(3)}_R$ flips its sign, leading to destructive inteference between PM-THG and electronic-THG polarization fields. Therefore, for the center frequency of pump \ding{173} in Fig.~\ref{detune_pump} (marked by the yellow dashed line \ding{173} in Fig.~\ref{fig:susceptibility}), the total THG intensity is suppressed below the pure electronic THG, as seen in Fig.~3(d) of the main text. Notably, the THG signal cannot be completely eliminated even with a perfect cancellation between $\chi^{(3)}_e$ and the real part of $\chi^{(3)}_R$ , as a result of the minor contribution from the imaginary part of $\chi^{(3)}_R$.
This explains why the minimum enhancement factor, illustrated in Fig.~4(b) of main text, is not strictly zero but rather on the order of $10^{-4}$.  

We note that Ref.~\cite{kinder2021detection} detected an asymmetric THG response in HCl vapor, attributed to interference between THG from the sample and the background THG from gas vessel windows. Our findings suggest this may instead be due to the intrinsic interference effect we have observed, which could not be disentangled by adjusting the pump polarization direction in the randomly oriented gas.

\bibliography{supplement}